\documentclass[aps,amsmath,amssymb,floatfix,onecolumn]{revtex4-1}
\usepackage{graphicx}
\usepackage{amsmath}
\usepackage{epstopdf}
\usepackage{amssymb}
\usepackage{amsthm}
\usepackage{color}
\usepackage{placeins}
\usepackage{subfigure}
\usepackage{setspace}
\usepackage{array}

\begin{document}

\title{Boundary layers in rotating weakly turbulent Rayleigh-B\'enard convection.}

\author{Richard J.A.M. Stevens$^1$}
\author{Herman J.H. Clercx$^{2,3}$}
\author{Detlef Lohse$^1$}
\affiliation{$^1$Department of Science and Technology and J.M. Burgers Center for Fluid Dynamics, University of Twente, P.O Box 217, 7500 AE Enschede, The Netherlands}
\affiliation{$^2$Department of Applied Mathematics, University of Twente, P.O Box 217, 7500 AE Enschede, The Netherlands}
\affiliation{$^3$Department of Physics and J.M. Burgers Centre for Fluid Dynamics, Eindhoven University of Technology, P.O. Box 513, 5600 MB Eindhoven, The Netherlands}
\date{\today}



\begin{abstract}
The effect of rotation on the boundary layers (BLs) in a Rayleigh-B\'enard (RB) system at a relatively low Rayleigh number, i.e. $Ra = 4\times10^7$, is studied for different $Pr$ by direct numerical simulations and the results are compared with laminar BL theory. In this regime we find a smooth onset of the heat transfer enhancement as function of increasing rotation rate. We study this regime in detail and introduce a model based on the Grossmann-Lohse theory to describe the heat transfer enhancement as function of the rotation rate for this relatively low Ra number regime and weak background rotation $Ro\gtrsim 1$. The smooth onset of heat transfer enhancement observed here is in contrast to the sharp onset observed at larger $Ra \gtrsim 10^8$ by Stevens {\it{et al.}} [Phys. Rev. Lett. {\bf{103}}, 024503, 2009], although only a small shift in the $Ra-Ro-Pr$ phase space is involved.
\end{abstract}

\maketitle

\section{Introduction}

\begin{figure} [t]
\subfigure{\includegraphics[width=2.9in]{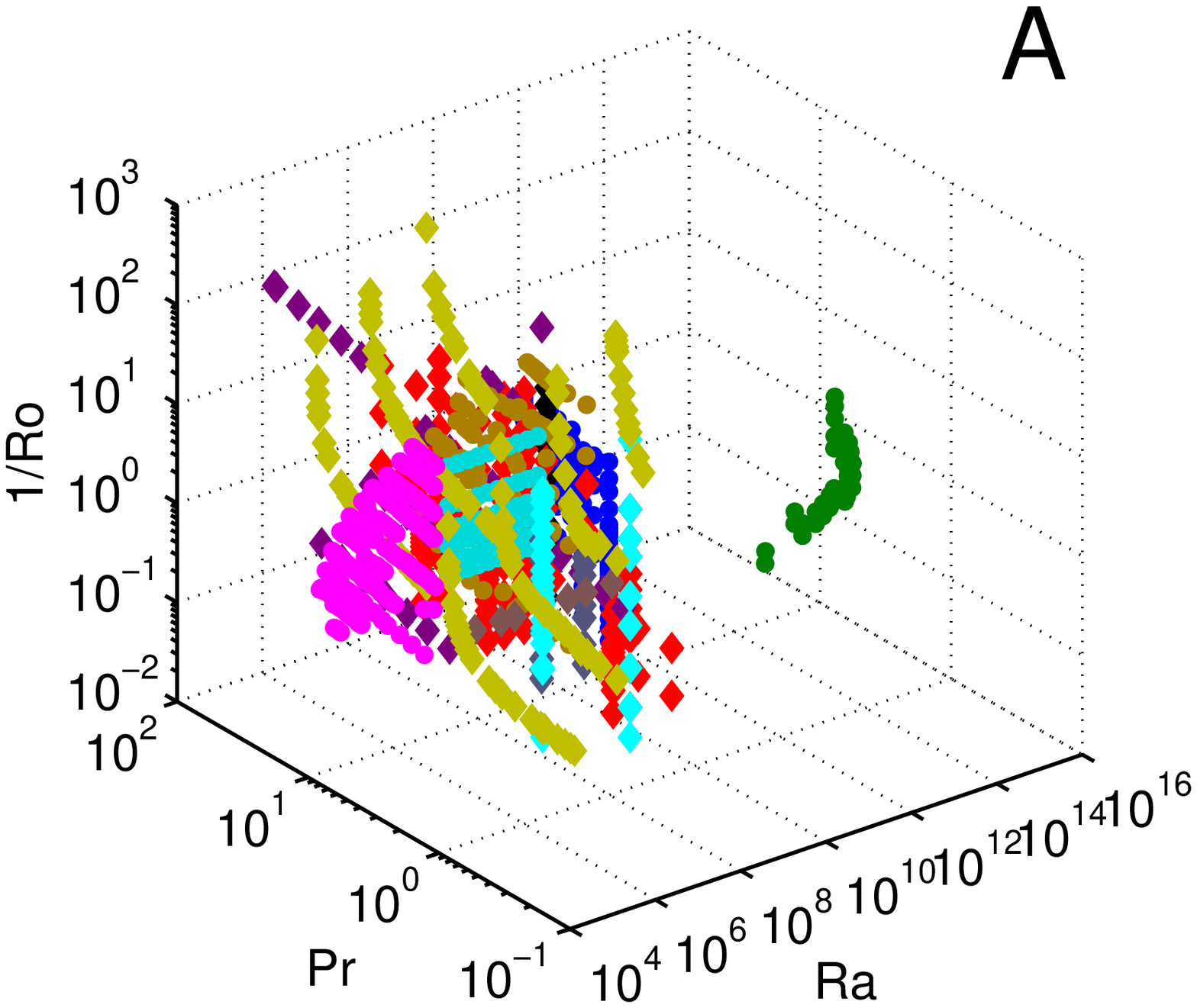}}
\subfigure{\includegraphics[width=2.9in]{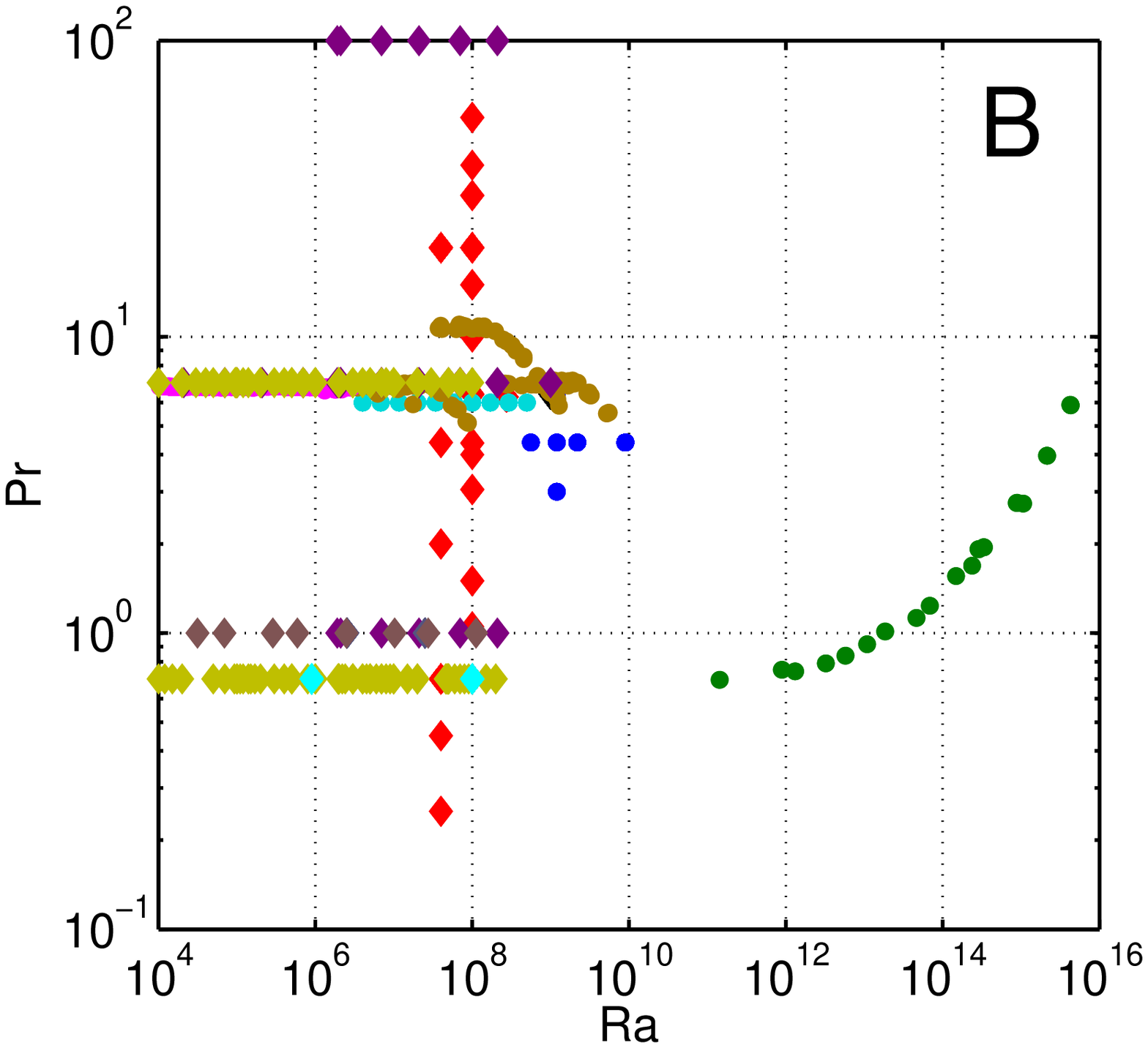}}
\subfigure{\includegraphics[width=2.9in]{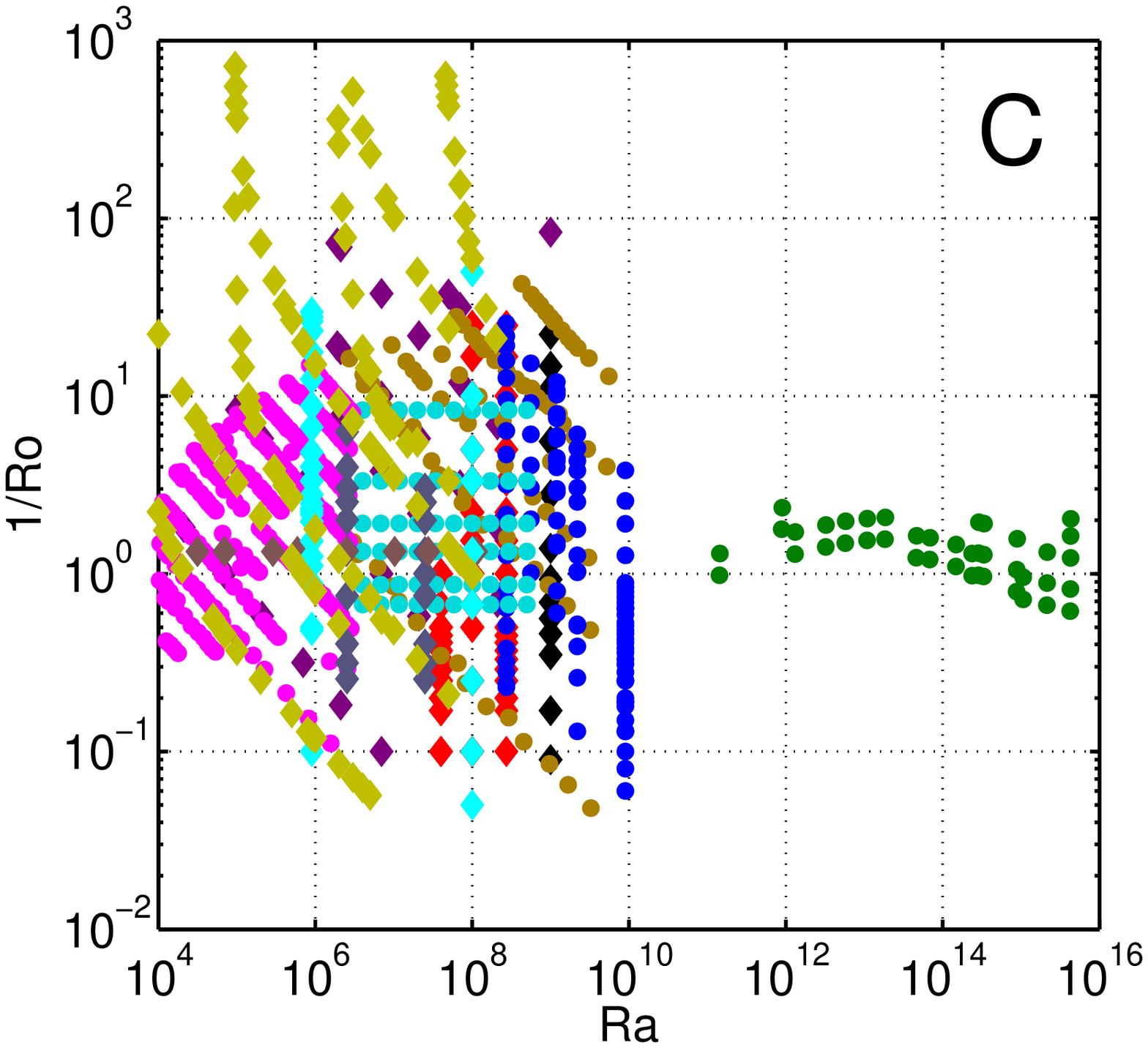}}
\subfigure{\includegraphics[width=2.9in]{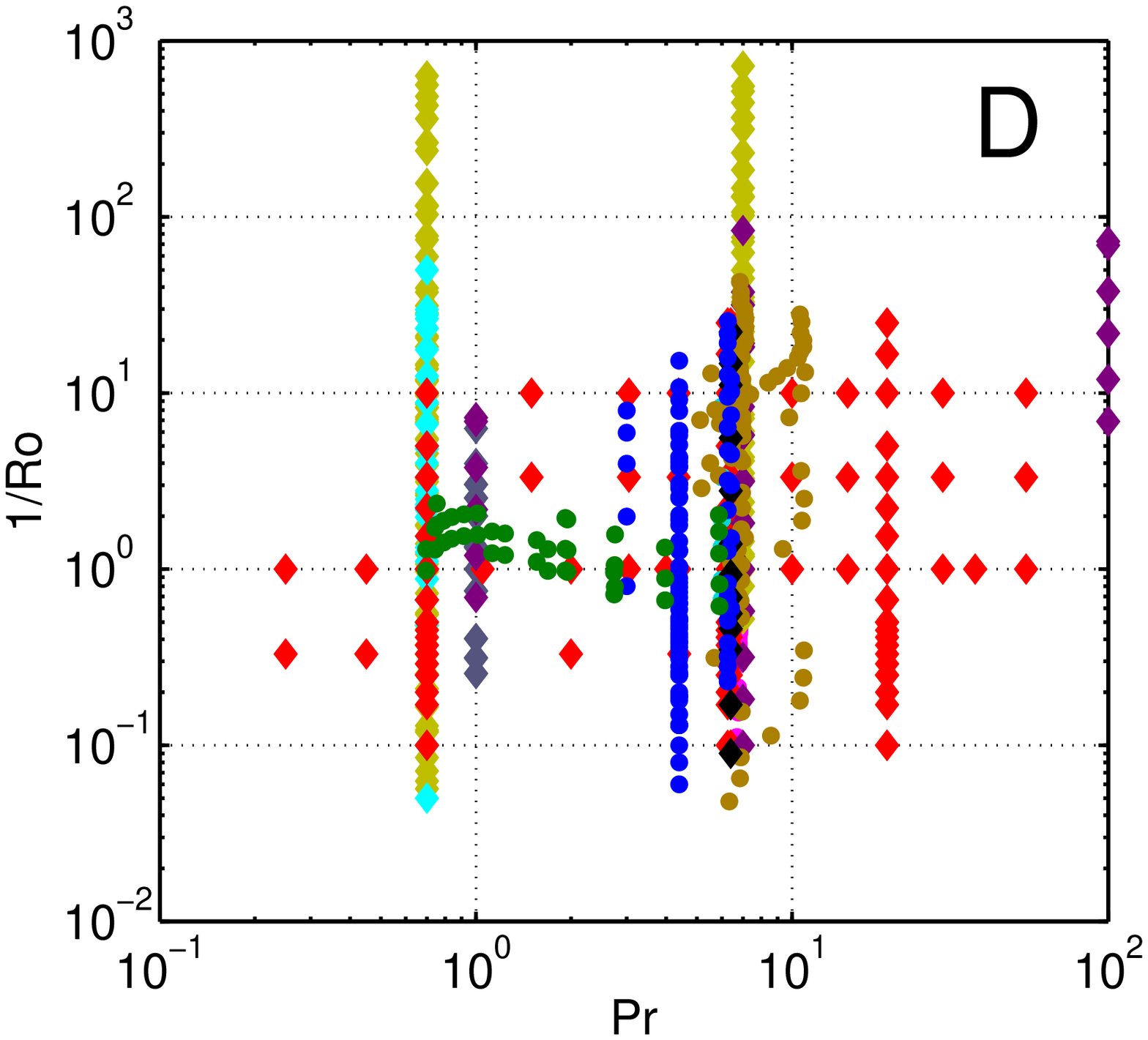}}
\subfigure{\includegraphics[width=2.0in]{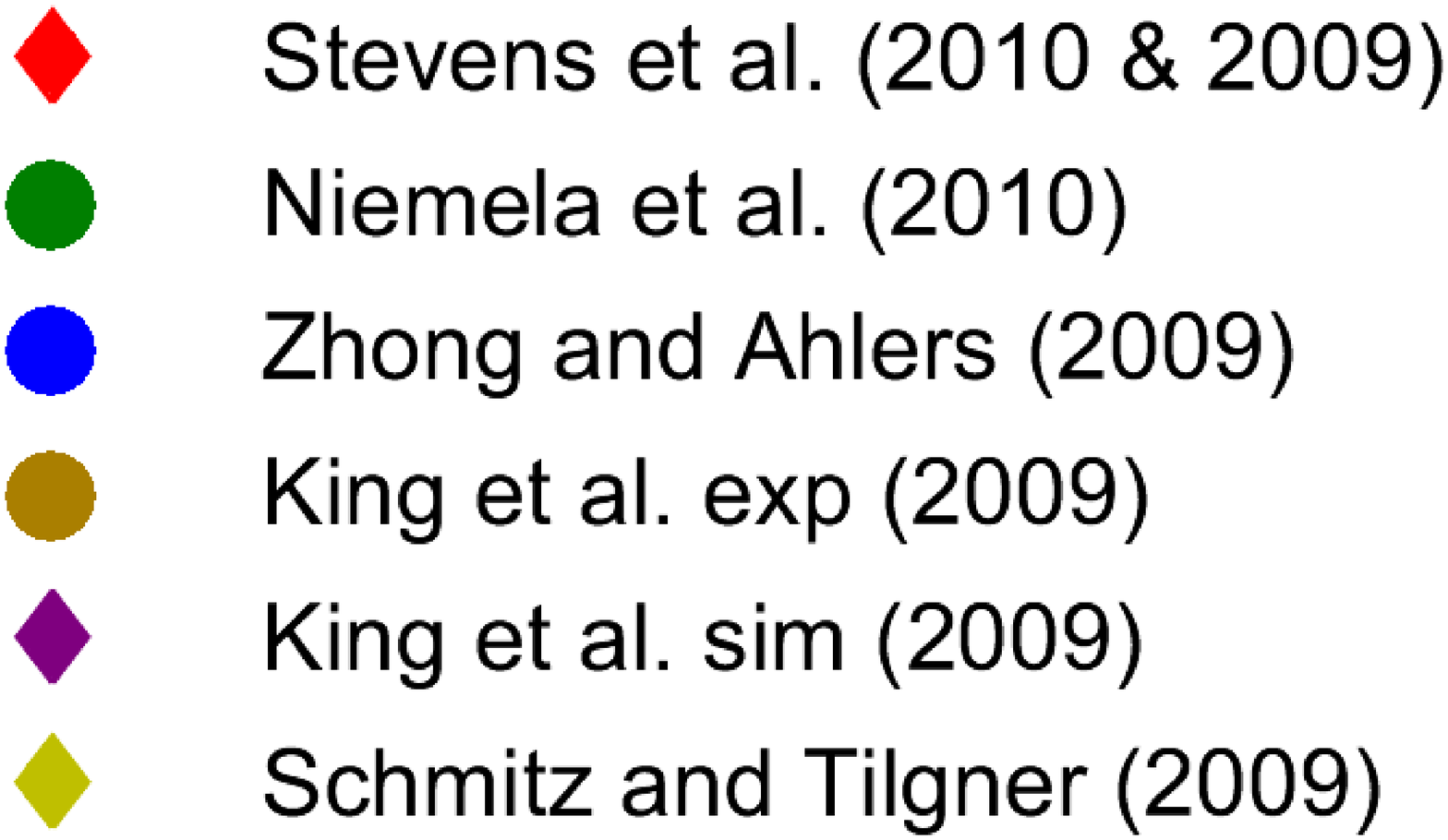}}
\subfigure{\includegraphics[width=2.0in]{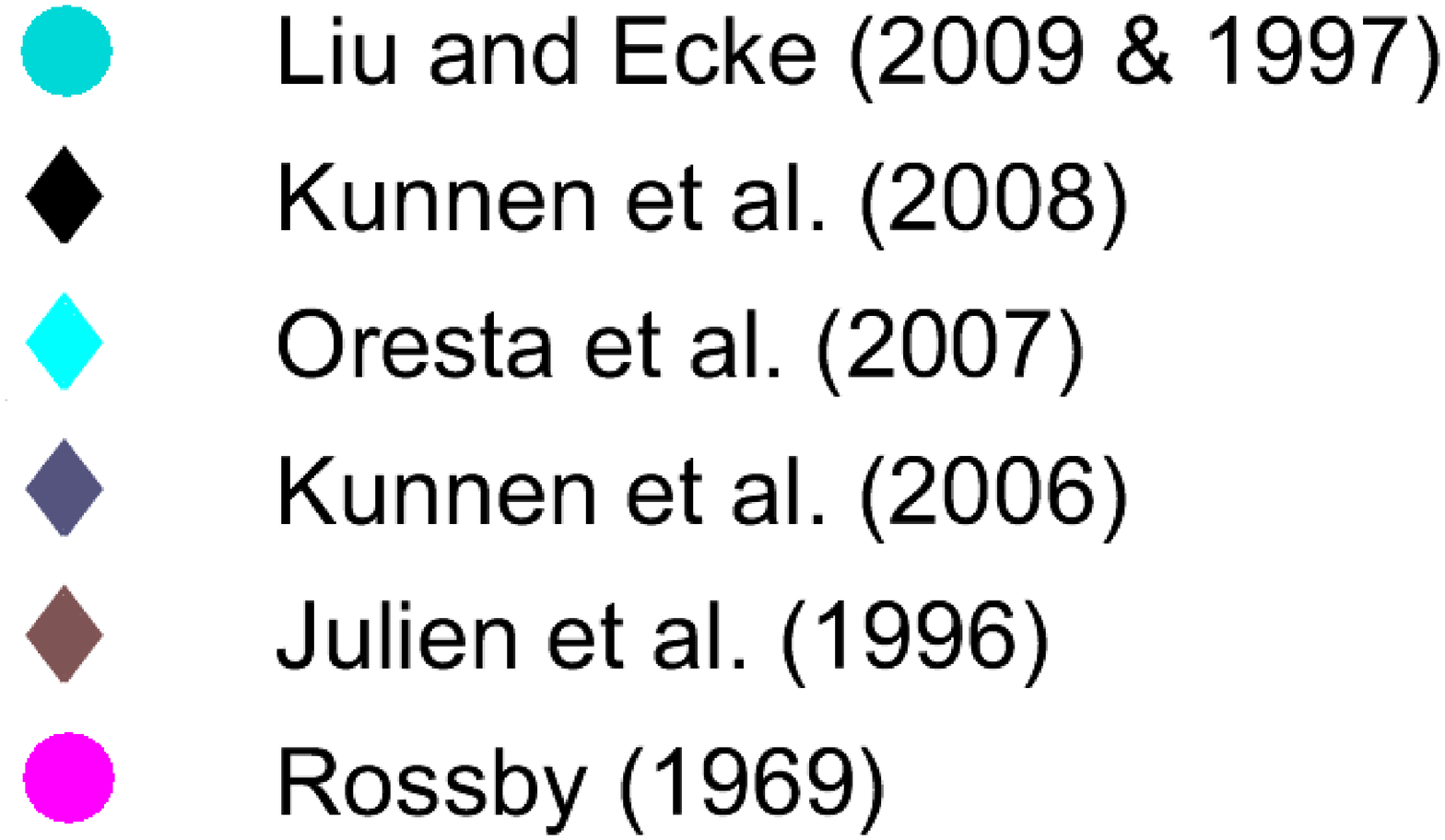}}
\caption{
{\small{
{\linespread{1.0}{
Phase diagram in Ra-Pr-Ro space for RRB convection. The data points indicate where $Nu$ has been measured or
 numerically calculated. The data are obtained in a cylindrical cell with aspect ratio $\Gamma=1$ with no slip boundary conditions, unless mentioned otherwise. The data from direct numerical simulations and experiments are indicated by diamonds and dots, respectively. The data sets, which are ordered chronologically in panel e, are from:
 Stevens et al. (2010 $\&$ 2009) \cite{zho09b,ste09,ste09d} and the simulations of this study;
 Niemela et al. (2010) \cite{nie10} ($\Gamma=0.5$);
 Zhong and Ahlers (2009) \cite{zho09b,ste09};
 King et al. (2009) \cite{kin09};
 Schmitz and Tilgner (2009) \cite{sch09} (free slip boundary conditions and horizontally periodic);
 Liu and Ecke (2009 $\&$ 1997) \cite{liu09,liu97} (square with $\Gamma=0.78$);
 Kunnen et al. (2008) \cite{kun08b};
 Oresta et al. (2007) \cite{ore07} ($\Gamma=0.5$);
 Kunnen et al. (2006) \cite{kun06} ($\Gamma=2$, horizontally periodic);
 Julien et al. (1996) \cite{jul96} ($\Gamma=2$, horizontally periodic);
 Rossby (1969) (varying aspect ratio).
 Panel a shows a three dimensional view on the phase space (see also the movie in the supplementary material),
 b) projection on the Ra-Pr phase space,
 c) projection on the Ra-Ro phase space,
 d) projection on the Pr-Ro phase space.
 }}
 }}
 }
\label{fig:Phase space}
\end{figure}

Normally the transition between different turbulent states is smooth, because the large random fluctuations that characterize the turbulent flow make sure that the entire phase space is explored and therefore the transitions between different states, that are explored as a control parameter is changed, are washed out. A classical system to study turbulence is Rayleigh B\'enard (RB) convection \cite{ahl09,ahl09b,loh10}. For given aspect ratio $\Gamma\equiv D/L$ ($D$ is the cell diameter and $L$ its height) and given geometry, its dynamics are determined by the Rayleigh number $Ra=\beta g\Delta L^3 /(\kappa \nu)$ and the Prandtl number $Pr=\nu/\kappa$. Here $\beta$ is the thermal expansion coefficient, $g$ the gravitational acceleration, $\Delta$ the temperature difference between the plates, and $\nu$ and $\kappa$ are the kinematic and thermal diffusivity, respectively. The heat transfer in a RB system can satisfactory be described by the Grossmann-Lohse (GL) theory \cite{ahl09,gro00,gro01,gro02,gro04} and shows that RB convection has different turbulent regimes in the $Ra-Pr$ phase space (see Fig. 3 of Ref.\ \cite{ahl09}). The case where the RB system is rotated around a vertical axis, i.e. Rotating Rayleigh-B\'enard (RRB) convection, at an angular speed $\Omega$ is interesting for industrial applications and problems in geology, oceanography, climatology, and astronomy. The rotation rate of the system is non-dimensionalized in the form of the Rossby number $Ro=\sqrt{\beta g \Delta/L}/(2\Omega)$. The dynamics of RRB convection are thus determined by three control parameters, i.e. $Ra$, $Pr$, and $Ro$, and this leads to a huge $Ra-Pr-Ro$ phase space, see Fig. \ref{fig:Phase space}.

\begin{figure}
  \centering
  \subfigure[]{\includegraphics[width=0.48\textwidth]{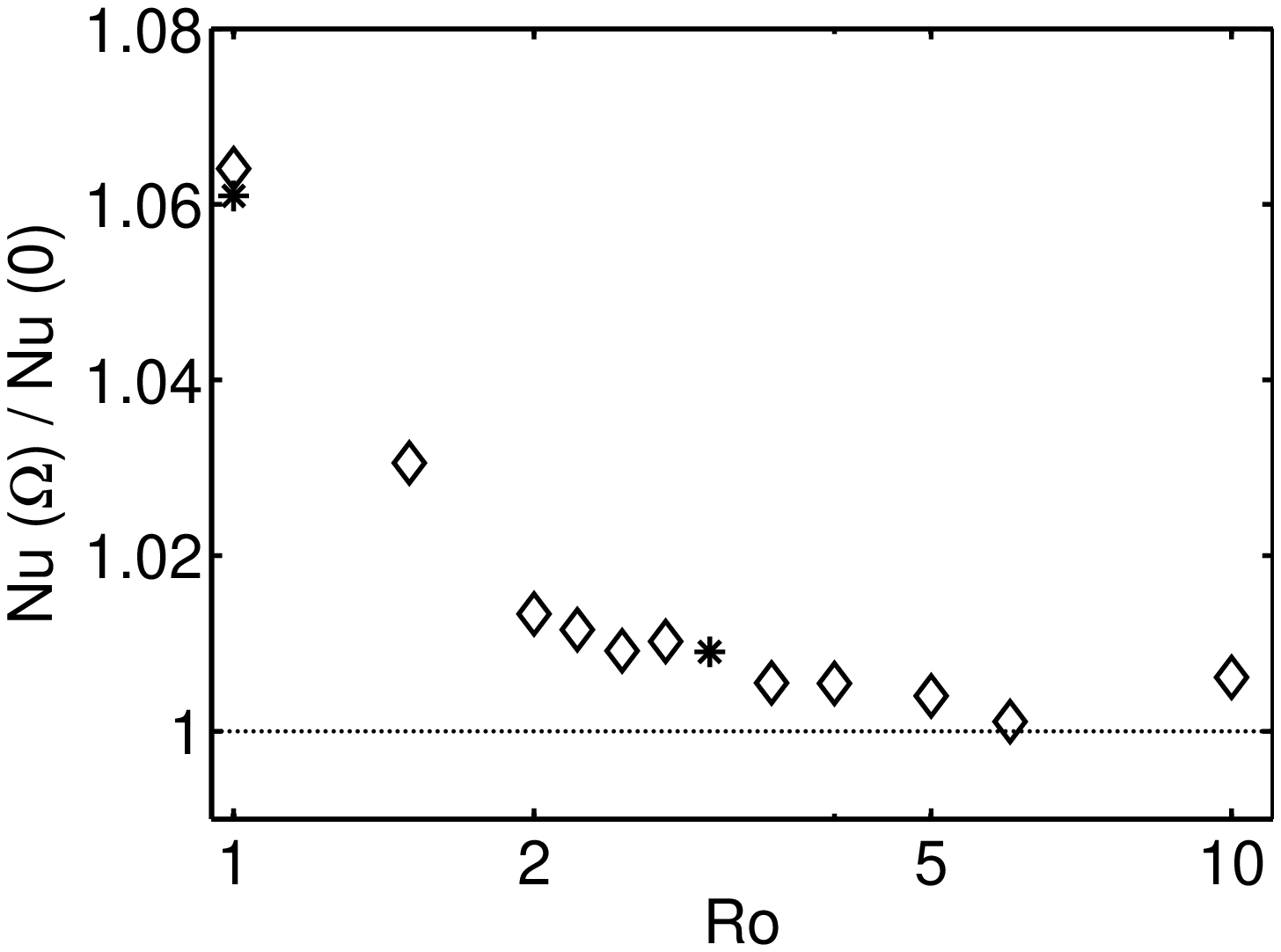}}
  \subfigure[]{\includegraphics[width=0.48\textwidth]{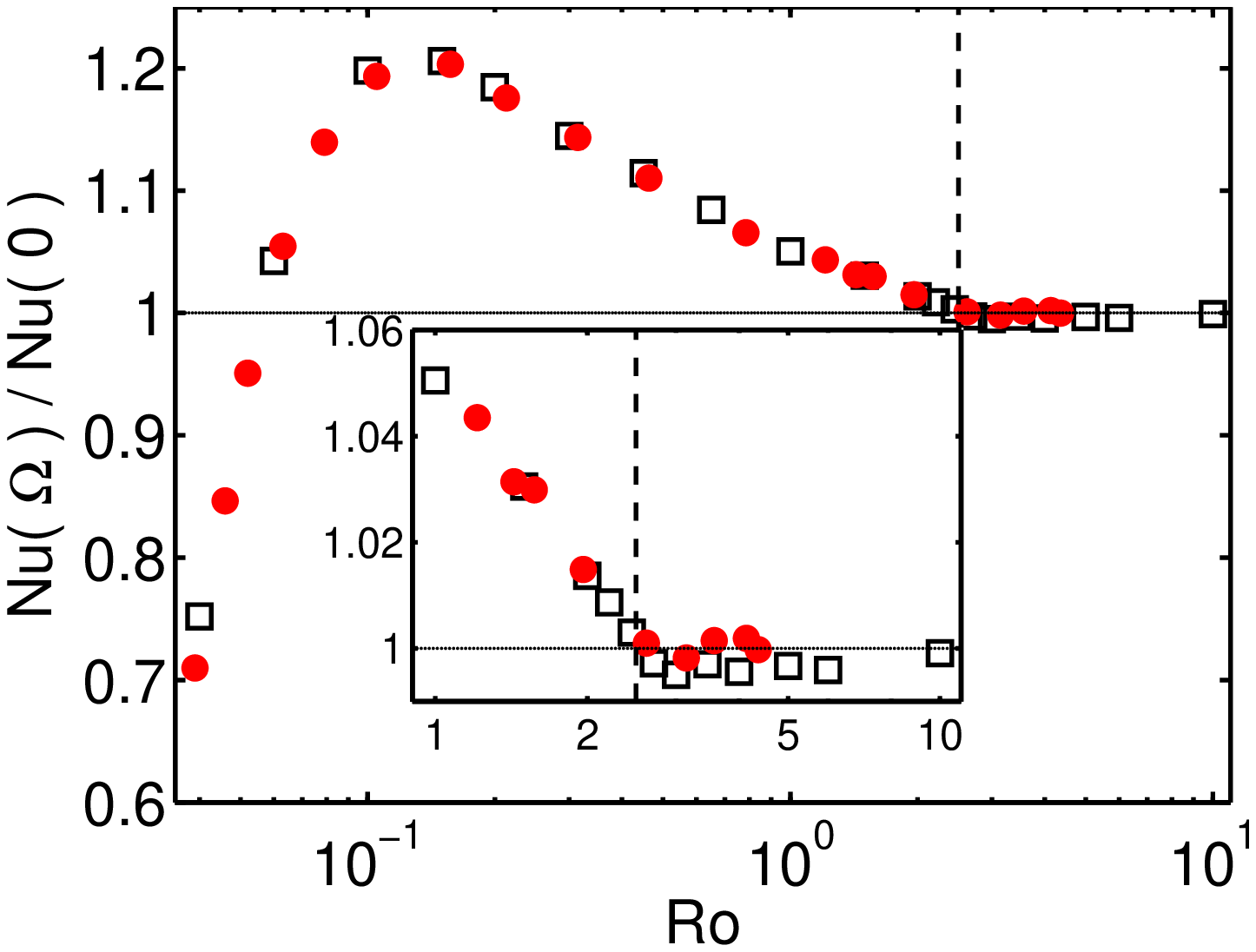}}
  \caption{Normalized heat transfer as a function of the Rossby number. a) Numerical results for $Ra=4\times10^7$ and $Pr=6.4$. The open black squares and stars indicate the results obtained on the $193 \times 65 \times 129$ and $257 \times 97 \times 193$ grid, respectively. b) $Ra=2.73\times10^8$ and $Pr=6.26$. Red solid circles: experimental data \cite{ste09}. Open black squares: numerical results \cite{zho09b,ste09}. The vertical dashed line indicates the position of the onset.}
  \label{Fig_SCL09_highRa_a}
\end{figure}

It is widely understood \cite{cha81} that rotation suppresses convective flow, and with it convective heat transport, when the rate of rotation is sufficiently large. However, experimental \cite{liu97,liu09,ros69,pfo84,bou86,bou90,zho93,kin09,zho09b,ste09} and numerical \cite{ore07,jul96,spr06,kun08b,kin09,sch09,zho09b,ste09,ste09d} studies on RRB convection have shown that rotation can also enhance the heat transport with respect to the non-rotating case. This heat transport enhancement is caused by Ekman pumping \cite{ros69,zho93,jul96,vor98,vor02,kun08b,kin09,zho09b,ste09,ste09d} and its efficiency depends strongly on the combination of $Ra$, $Pr$, and $Ro$ \cite{zho09b,ste09,ste09d}. In this paper we will discuss the results of Direct Numerical Simulations (DNS) that show that this heat transfer enhancement as function of the Ro number is smooth for relatively low Ra number, here  $Ra=4\times 10^7$ (see Fig. \ref{Fig_SCL09_highRa_a}a), while experimental and numerical data for $Ra=2.73 \times 10^8$ and $Pr=6.26$ show a sharp onset for the heat transport enhancement, see Fig. \ref{Fig_SCL09_highRa_a}b. This difference is remarkable since only a small shift in the $Ra-Pr-Ro$ phase space is involved (see Fig. \ref{fig:Phase space}).

In this paper we will first describe the flow characteristics found in the simulations. We will show that there is a smooth transition from one turbulent regime to another for the relatively low Ra number regime whereas a sharp transition is found for higher $Ra$. In section \ref{Sec2} we will discuss the properties of the BLs found in the DNS in detail. Subsequently the laminar BL theory for flow over an infinitely large rotating disk will be discussed in section \ref{Sec3} in order to explain the BL properties found in the DNS. The derived scaling laws from this theory will be used in a model based on the GL theory to describe the heat transfer enhancement as function of $Ro$ for the relatively low $Ra$ number regime with weak background rotation, see section \ref{Sec4}.

\section{Numerical results for boundary layers in RRB convection}
\label{Sec2}
The flow characteristics of RRB convection for $Ra=4\times 10^7$, $1<Ro<\infty$, and $0.2<Pr<20$, are obtained from solving the three-dimensional Navier-Stokes equations within the Boussinesq approximation:

\begin{eqnarray}
 \frac{D\textbf{u}}{Dt} &=& - \nabla P + \left( \frac{Pr}{Ra} \right)^{1/2} \nabla^2 \textbf{u} + \theta \textbf{$\widehat{z}$}- \frac{1}{Ro} \widehat{z} \times \textbf{u}, \\
 \frac{D\theta}{Dt} &=& \frac{1}{(PrRa)^{1/2}}\nabla^2 \theta ,
\end{eqnarray}
with
 $\nabla \cdot \textbf{u} = 0$.
 Here \textbf{$\widehat{z}$} is the unit vector pointing in the opposite direction to gravity,
 $D/Dt = \partial_t + \textbf{u} \cdot \nabla $ the material derivative,
 $\textbf{u}$ the velocity vector (with no-slip boundary conditions
 at all walls), and $\theta$ the non-dimensional temperature, $0\leq \theta \leq 1$.
 Finally, $P$ is the reduced pressure (separated from its hydrostatic contribution, but containing the centripetal contributions): $P=p - r^2/(8Ro^2)$, with $r$ the distance to the rotation axis. The equations have been made non-dimensional by using, next to $L$ and $\Delta$,
 the free-fall
 velocity $U=\sqrt{\beta g \Delta L}$. A constant temperature boundary condition is used at the bottom and top plate and the side wall is adiabatic. Further details about the numerical procedure can be found in Refs.\ \cite{ver96,ver99,ver03}.

The first set of simulations is used to study the $Ro$ number dependence of the following quantities: the normalized heat transfer, the thickness of the thermal BL, and the normalized averaged root mean square (rms) vertical velocity fluctuations. Here we have simulated RRB convection at several $Ro$ numbers for three different $Pr$ numbers ($Pr=0.7$, $Pr=6.4$, and $Pr=20$). All these simulations are performed on a grid with $65 \times 193 \times 129$ nodes, respectively, in the radial, azimuthal, and vertical directions, allowing for sufficient resolution in the bulk and the BL according the resolution criteria set in Ref. \cite{ste09b}. The Nusselt number is calculated in several ways as is discussed in detail in Ref.~\cite{ste09b} and its statistical convergence has been controlled. Some data for $Pr=6.4$ have already been published in Ref. \cite{ste09}. There the average was over $4000$ dimensionless time units. The new results for $Pr=0.7$ and $Pr=20$ are averaged over $2500$ dimensionless time units. Note that we simulated the flow for a large number of eddy turnover times to reduce the statistical error in the obtained Nusselt number results and to prevent the influence of transient effects. This is necessary to accurately resolve the transition regime where the heat transfer starts to increase and to accurately determine the flow statistics. Furthermore, we note that all simulations are started from a new flow field in order to rule out hysteresis effects.

\begin{figure}
  \centering
  \subfigure[]{\includegraphics[width=0.48\textwidth]{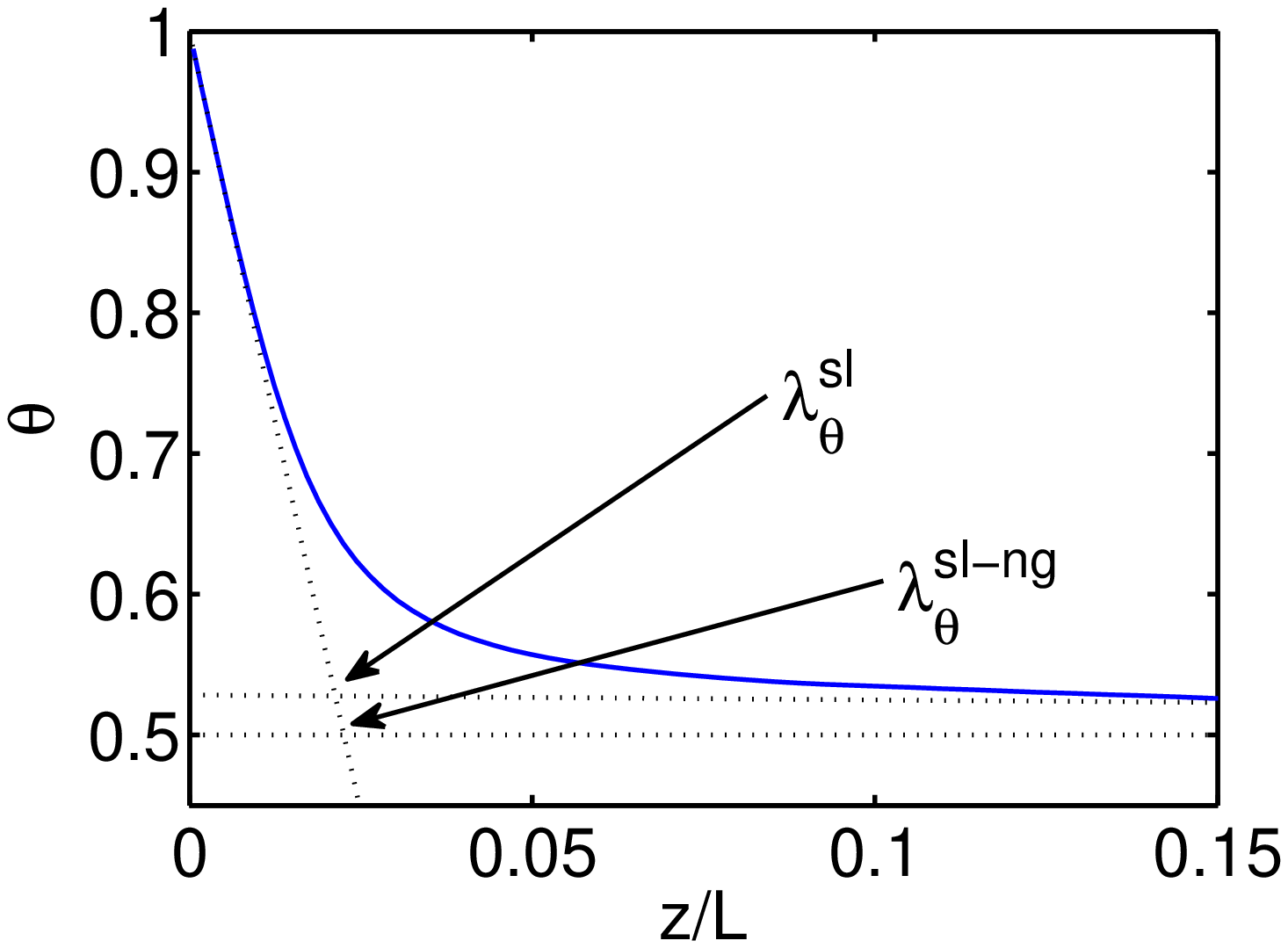}}
  \subfigure[]{\includegraphics[width=0.48\textwidth]{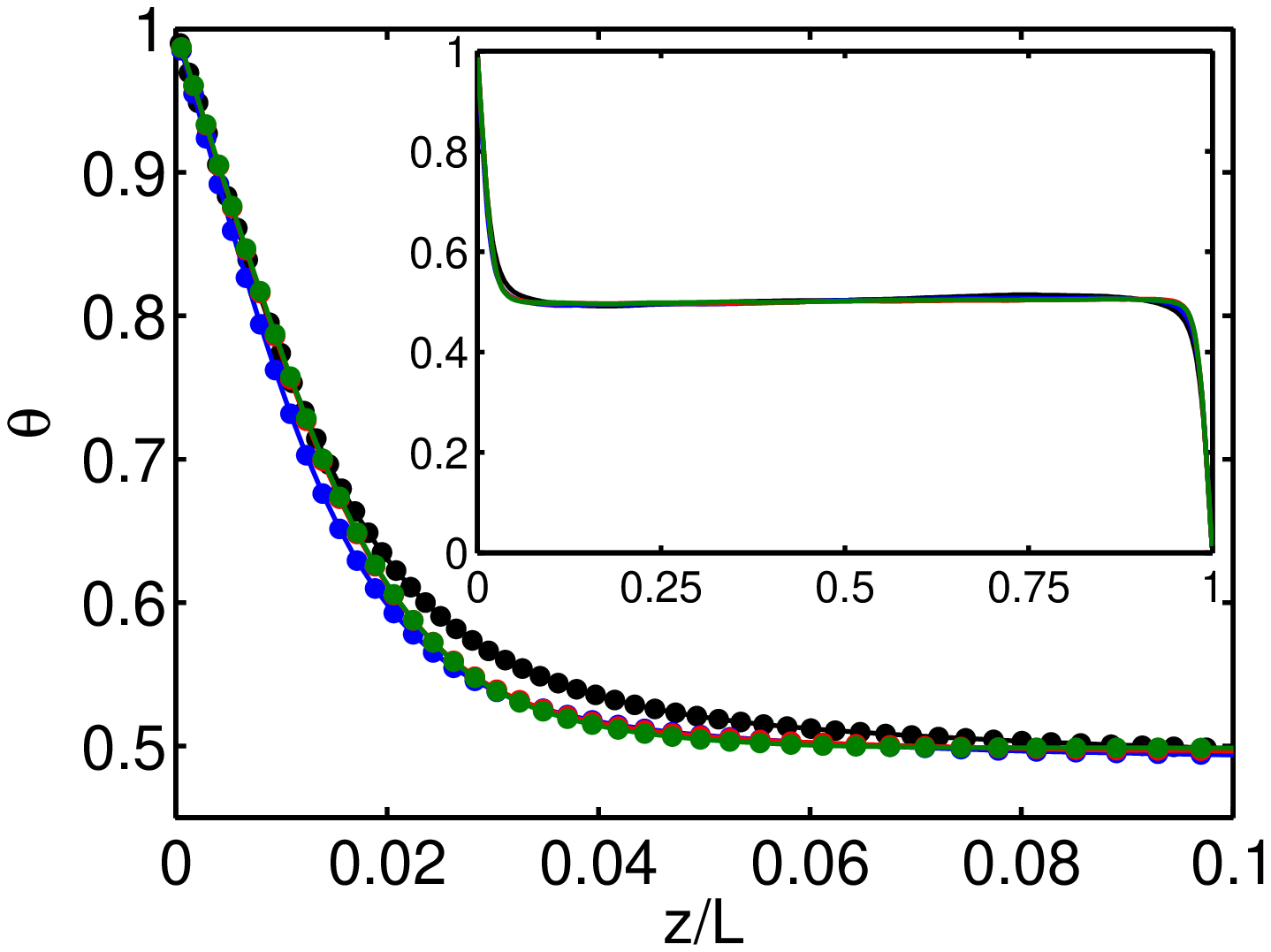}}
  \subfigure[]{\includegraphics[width=0.48\textwidth]{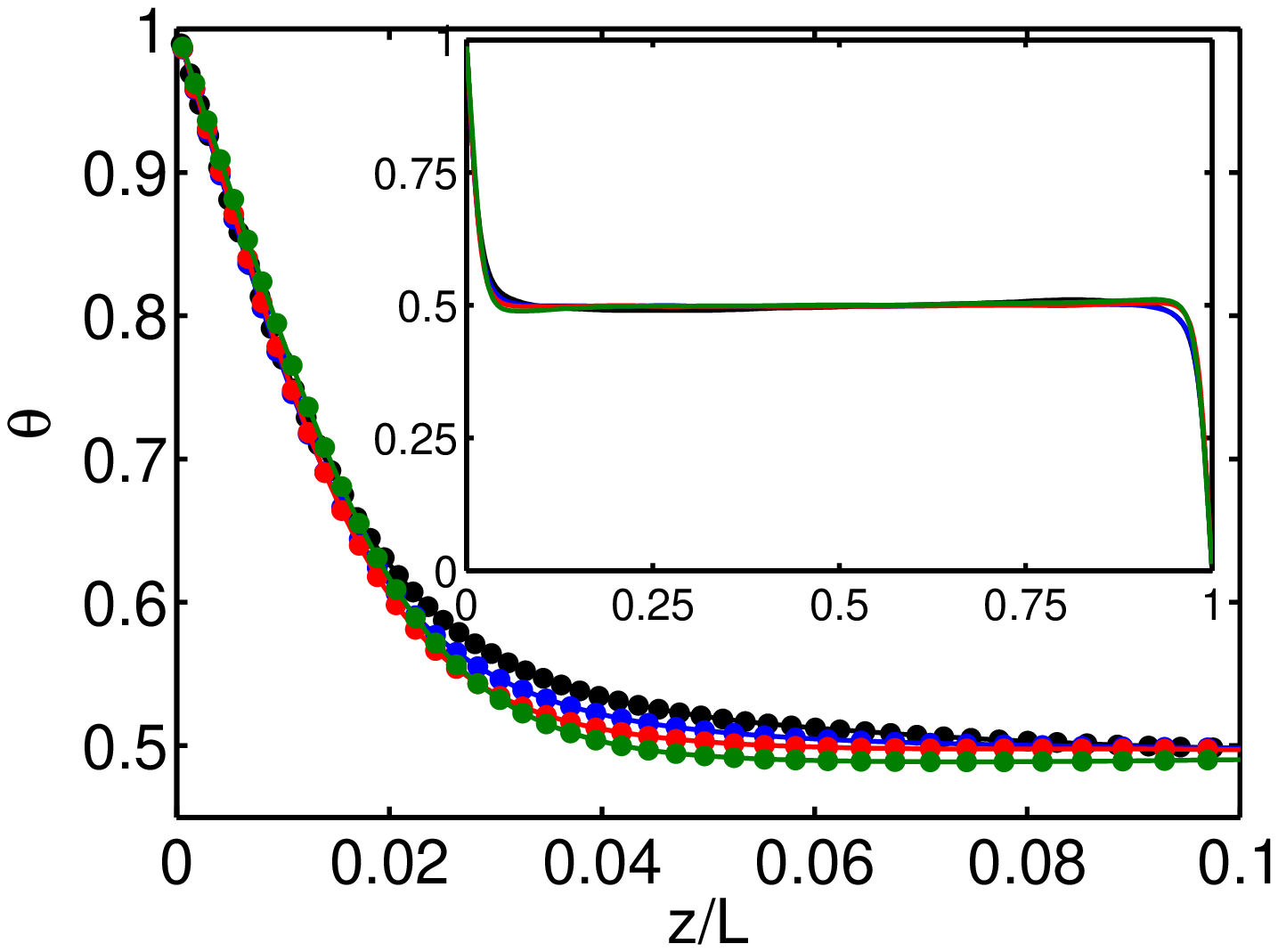}}
  \subfigure[]{\includegraphics[width=0.48\textwidth]{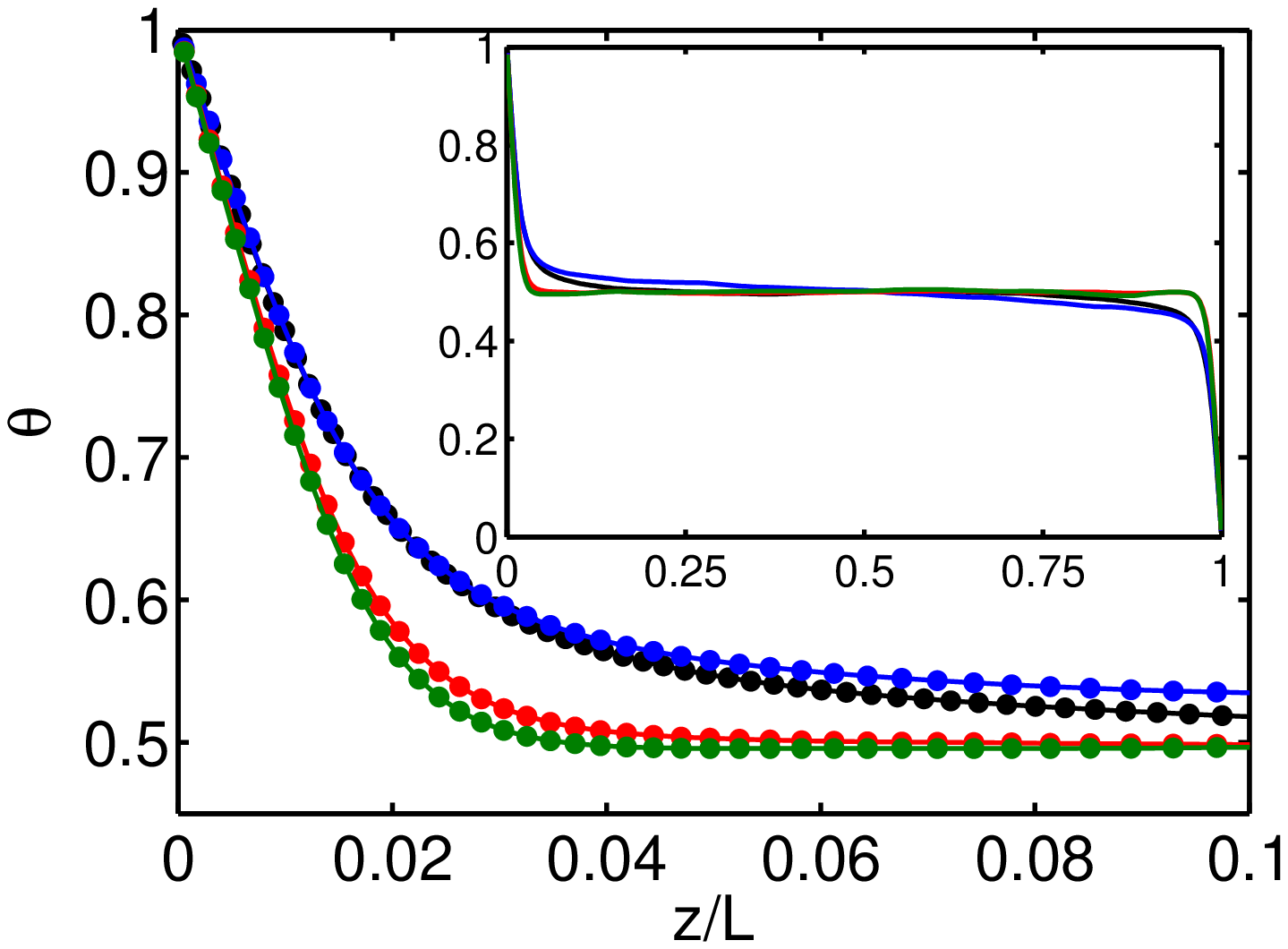}}
  \caption{a) Visualization of the definitions of the thermal BL thicknesses $\lambda_\theta^{sl}$ and $\lambda_\theta^{sl-ng}$ (assuming zero gradient in the bulk). b-d) Azimuthally averaged temperature profiles for $Ra=4\times10^7$ and different $Pr$ and $Ro$ numbers at the cylinder axis $r=0$ for b) $Ro=\infty$, c) $Ro=3$, and d) $Ro=1$. Black, blue, red, and dark green indicate the profiles for $Pr=0.25$, $Pr=0.7$, $Pr=6.4$, and $Pr=20$, respectively. The dots indicate the data points obtained from the simulations. The insets show the profile over the full cell.}
  \label{Fig_SCL09_profile_temp}
\end{figure}

\begin{figure}
  \centering
  \subfigure[]{\includegraphics[width=0.48\textwidth]{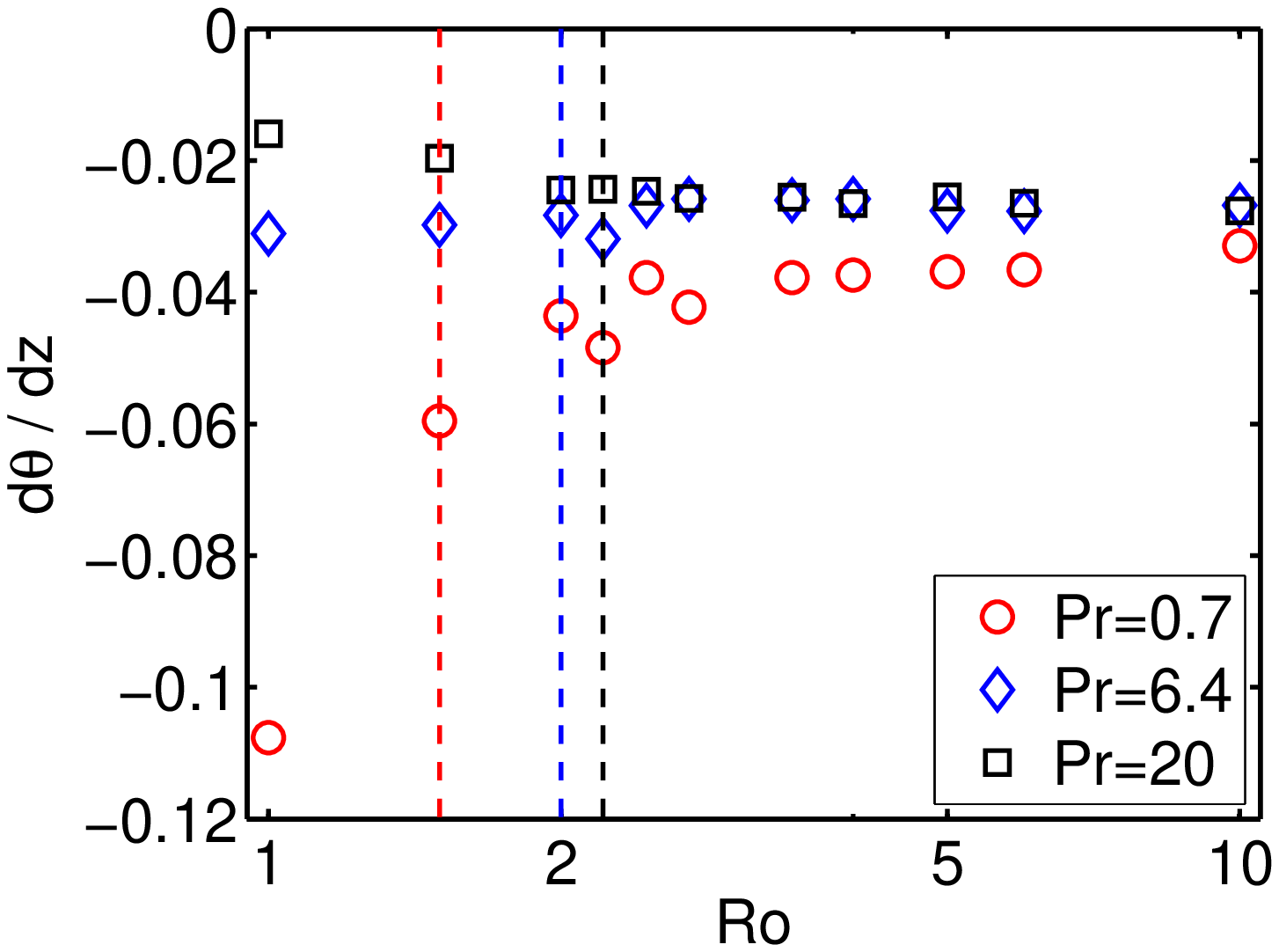}}
  \subfigure[]{\includegraphics[width=0.48\textwidth]{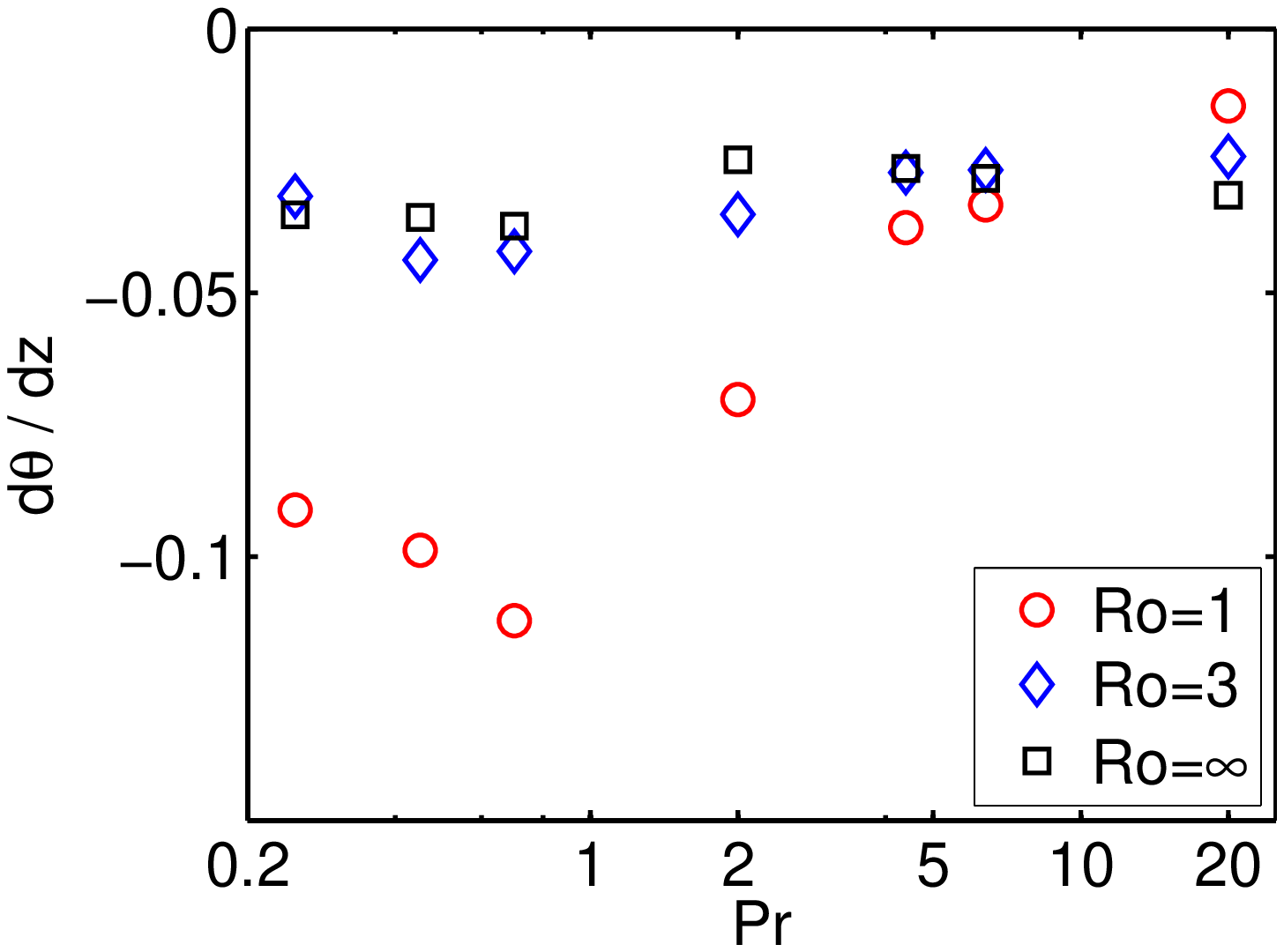}}
  \caption{a) Horizontally averaged temperature gradient at midheight as a function of $Ro$ for $Ra=4\times10^7$ and different $Pr$. Red open circles, blue diamonds, and black squares are the data for $Pr=0.7$, $Pr=6.4$, and $Pr=20$, respectively. b) Temperature gradient at midheight as function of $Pr$ for $Ra=4\times10^7$ and different $Ro$. Red open circles, blue diamonds, and black squares are the data for $Ro=1$, $Ro=3$, and $Ro=\infty$, respectively. The vertical dashed lines in both graphs represent the point at which the LSC strength starts to decrease, see Fig. \ref{Fig_SCL09_flow}.}
  \label{Fig_SCL09_gradients}
\end{figure}

\begin{figure}
  \centering
  \subfigure[]{\includegraphics[width=0.48\textwidth]{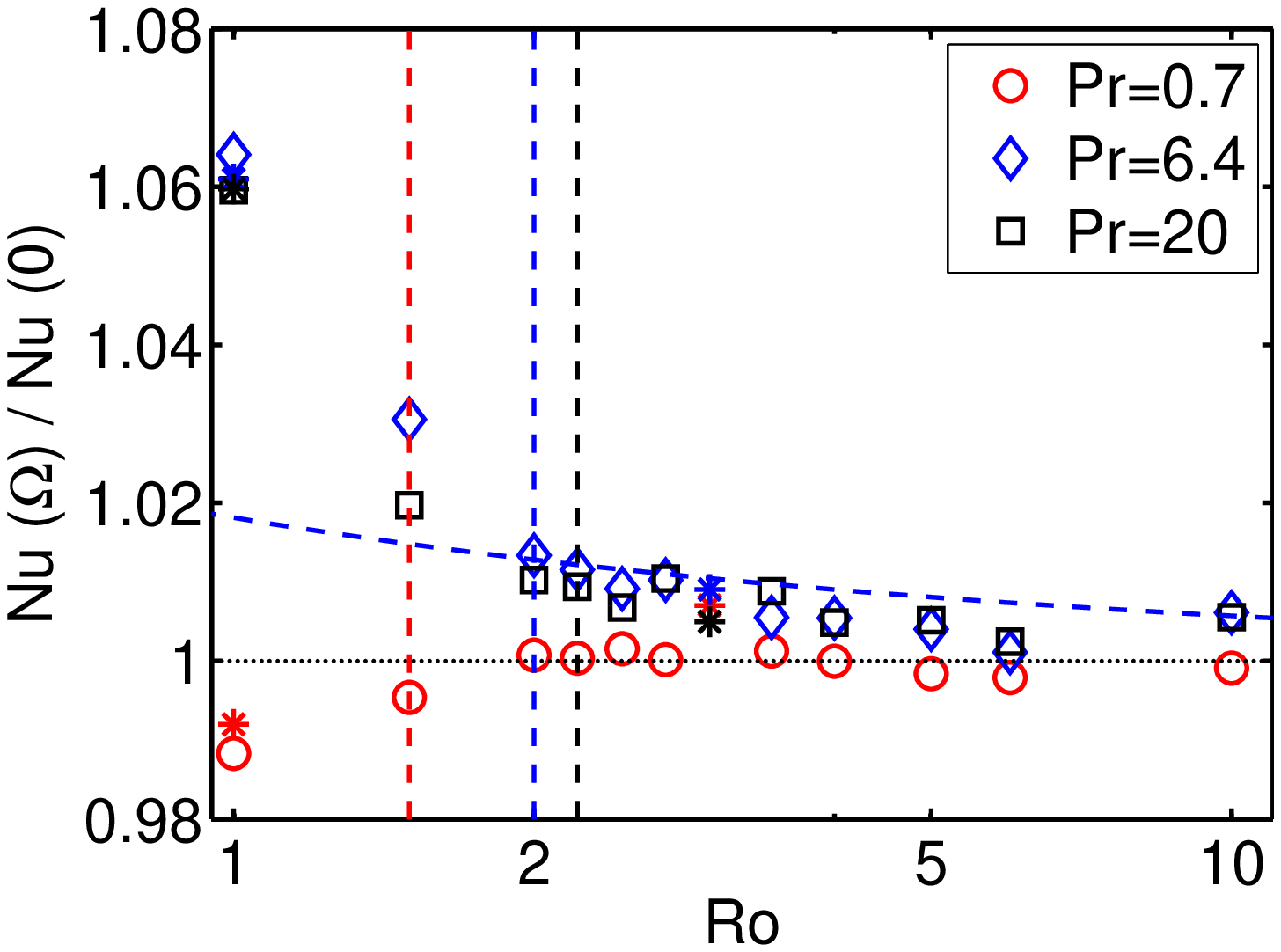}}
  \subfigure[]{\includegraphics[width=0.48\textwidth]{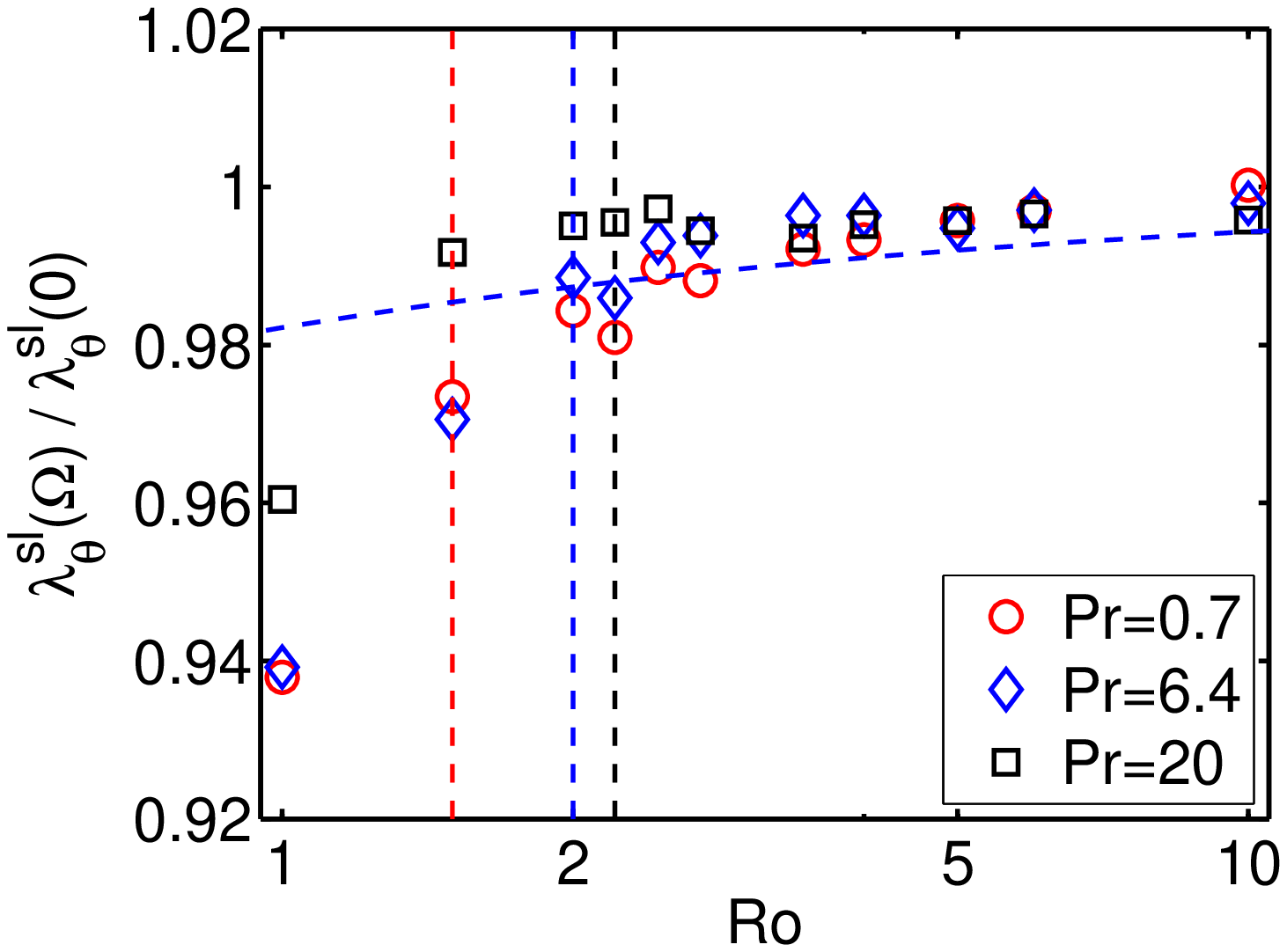}}
  \caption{a) Normalized heat transfer as a function of $Ro$ for $Ra=4\times10^7$ and different $Pr$. Red open circles, blue diamonds, and black squares are the data for $Pr=0.7$, $Pr=6.4$, and $Pr=20$, respectively. The dash-dotted line is the fit obtained by the model Eq.\ \ref{Eq model}, with $\alpha=55$. The stars indicate the results from the grid refinement check (see text).
  b) Horizontally averaged thickness of the thermal BL $\lambda^{sl}_{\theta}$ as function of $Ro$. Symbols as in Fig. \ref{Fig_SCL09_model}a and the dash-dotted line indicates the same model fit as in Fig. \ref{Fig_SCL09_model}a. The vertical dashed lines in both graphs represent the point at which the LSC strength starts to decrease, see Fig. \ref{Fig_SCL09_flow}.}
  \label{Fig_SCL09_model}
\end{figure}

The second set of simulations is used to study the $Pr$ number dependence of the same set of quantities. Here we simulated RRB for several $Pr$ numbers and three different $Ro$ numbers ($Ro=\infty$, $Ro=3$, and $Ro=1$). The simulations for $Pr\geqq 0.70$ are performed on a $97 \times 257\times193$ and the simulations at $Pr=0.25$ and $Pr=0.45$ are performed on a $129 \times 385 \times257$ grid. The finer resolution is needed here as the structure of the flow changes and the Reynolds number based on the LSC increases for lower $Pr$. For most cases the flow is simulated for $400$ dimensionless time units and $200$ dimensionless time units were simulated before data are collected to prevent any influence of transient effects \cite{ste09b}. For $Pr=2$ and $Pr=4.4$ we averaged over $1200$ dimensionless time units in order to obtain more accurate statistics on the velocity field. Note that the second set of simulations is partially overlapping with the first set of simulations. We find that results obtained on the different grids, are very similar, i.e. the difference is generally between $0.5\%$ and $1\%$, see Figs. \ref{Fig_SCL09_highRa_a}a, \ref{Fig_SCL09_model}a and \ref{Fig_SCL09_model_Pr}a. Again all simulations are started from a new flow field in order to rule out any hysteresis effects.

\begin{figure}
  \centering
  \subfigure[]{\includegraphics[width=0.32\textwidth]{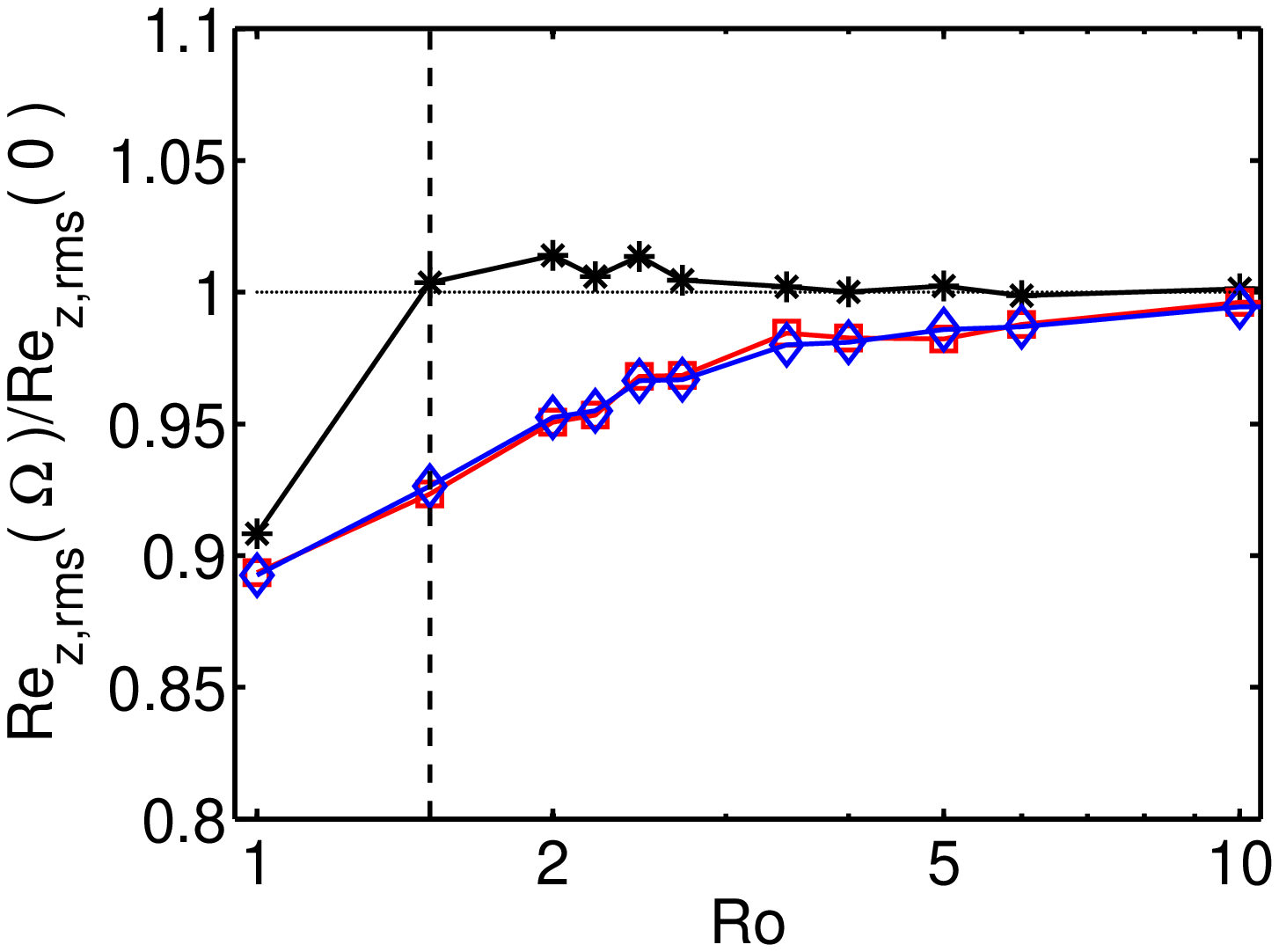}}
  \subfigure[]{\includegraphics[width=0.32\textwidth]{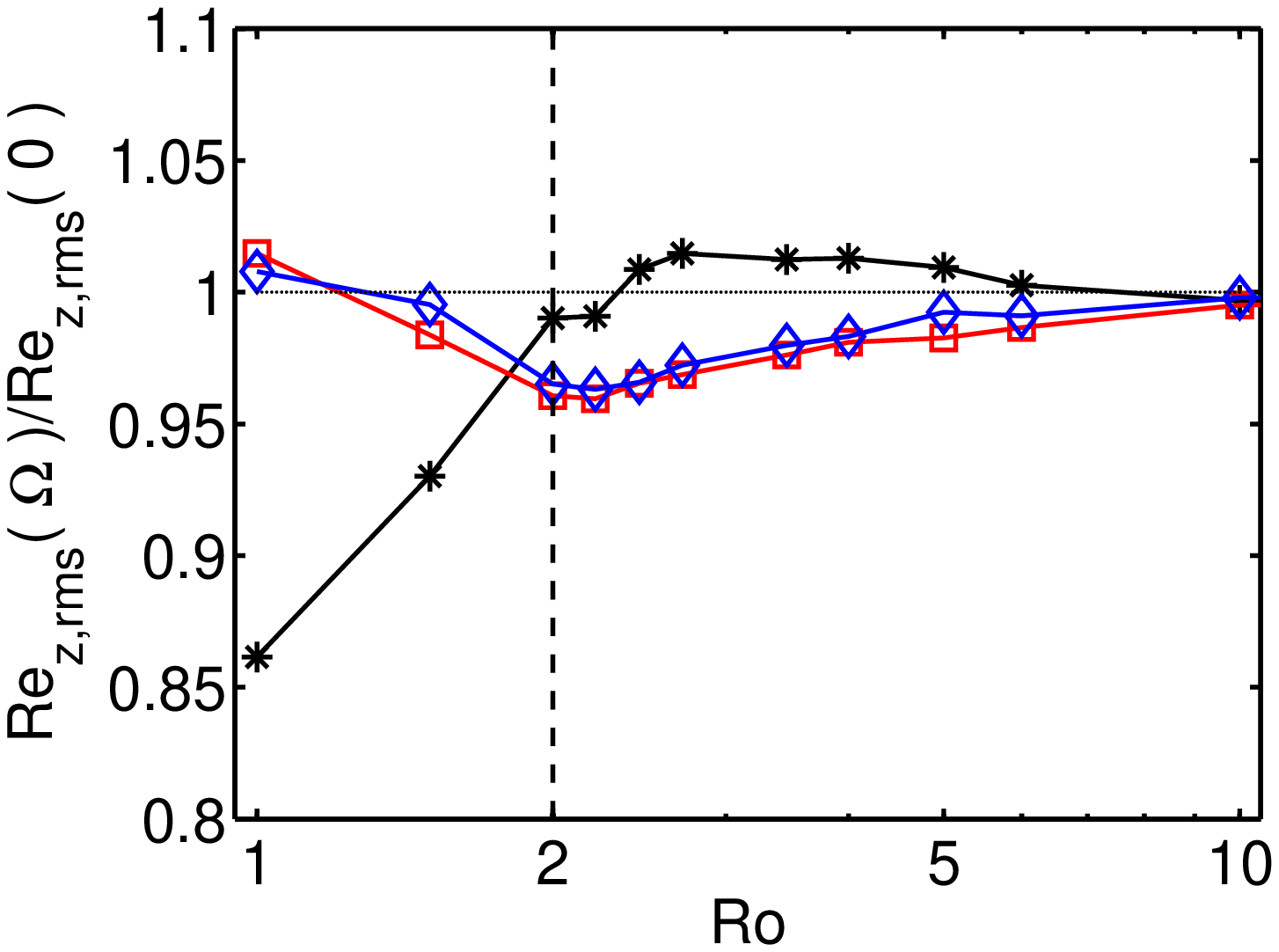}}
  \subfigure[]{\includegraphics[width=0.32\textwidth]{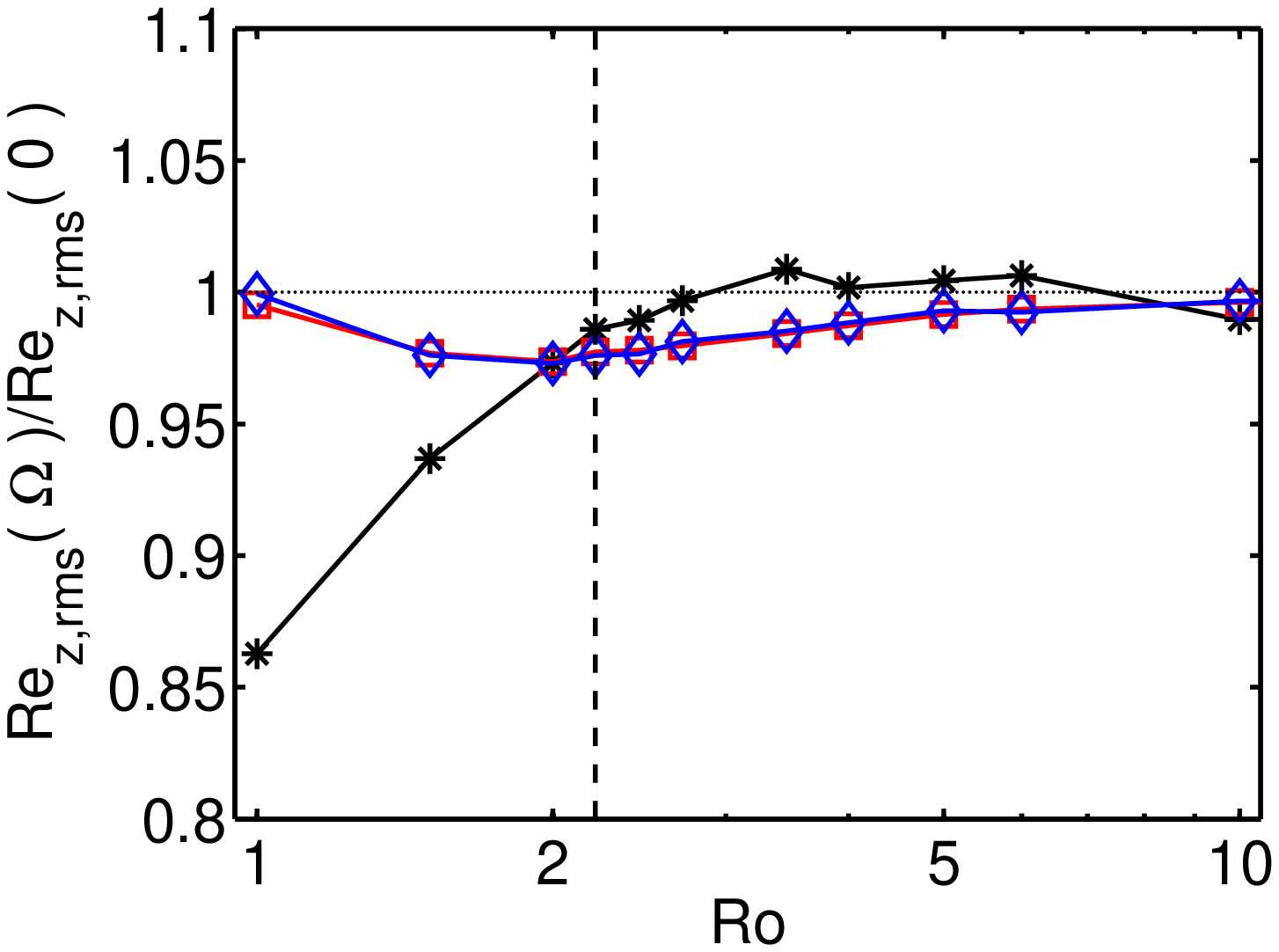}}
  \caption{The normalized averaged rms vertical velocity fluctuations $Re_{z,rms}$ for $Ra=4\times 10^7$ and different $Pr$ as function of $Ro$. The black line indicates the volume averaged value of $Re_{z,rms}$. The red and the blue line indicate the horizontally averaged values of the $Re_{z,rms}$ at a distance $\lambda_{\theta}^{sl}(r)$ from the lower and upper plate, respectively. The vertical dashed lines again indicate the position where the LSC strength starts to decrease. a) $Pr=0.7$, b) $Pr=6.4$, c) $Pr=20$.}
  \label{Fig_SCL09_flow}
\end{figure}

Fig. \ref{Fig_SCL09_profile_temp} shows the azimuthally averaged temperature profile at the cylinder axis for different system parameters. In previous numerical studies concerning (rotating) RB convection the thermal BL thickness is usually defined by either looking at the maximum rms value of the temperature fluctuations or by considering the BL thickness based on the slope of the mean temperature profile. In the latter case it is usually assumed that no mean temperature gradient exists in the bulk (the BL thickness according to this assumption is denoted by $\lambda^{sl-ng}_{\theta}$). The temperature gradient in the bulk is, however, strongly influenced by rotation \cite{zho09b,kun09}, and when rotation is present also by $Pr$, see Fig. \ref{Fig_SCL09_gradients}. We prefer to define the thermal BL thickness $\lambda_{\theta}^{sl}$ as the intersection point between the linear extrapolation of the temperature gradient at the plate with the behaviour found in the bulk, see Fig. \ref{Fig_SCL09_profile_temp}a. From now on this definition of the thermal BL thickness will be used here.

For the relatively low $Ra$ number regime, here $Ra=4\times10^7$, the heat transfer enhancement as function of $Ro$ is smooth, see Fig. \ref{Fig_SCL09_model}a. Note that although the behaviour of $\lambda_{\theta}^{sl}$ (see fig. \ref{Fig_SCL09_model}b), i.e. the horizontally averaged value of the radially dependent thermal BL thickness ($\lambda_\theta^{sl}(r)$), as function of $Ro$ is similar for all $Pr$ the behaviour of $Nu$ is very different. This is due to the influence of $Pr$ on the effect of Ekman pumping \cite{ste09d,zho09b}. At low $Pr$ the larger thermal diffusivity limits the effect of Ekman pumping and causes a larger destabilizing temperature gradient in the bulk \cite{zho09b,ste09d}, see Fig. \ref{Fig_SCL09_gradients}. Due to the limited effect of Ekman pumping there is no heat transport enhancement for low $Pr$, see Fig. \ref{Fig_SCL09_model}.

Fig. \ref{Fig_SCL09_flow} shows that the volume averaged $Re_{z,rms}$ (dimensionless rms velocity of the axial velocity fluctuations), which is a measure for the strength of the LSC, decreases strongly for strong enough rotation. The vertical dashed lines in Fig. \ref{Fig_SCL09_flow} indicate the position where $Re_{z,rms}(\Omega)$/$Re_{z,rms}(0)$ becomes smaller than $1$, which we use to indicate the point at which the LSC strength starts to decrease. This value is determined by extrapolating the behaviour observed at low $Ro$ numbers to reduce the effect of the uncertainty in single data points. In Ref. \cite{ste09} (see figure 3 of that paper) we also used this method to indicate the position of the onset of the heat transfer enhancement in the high $Ra$ number regime. However, for this lower $Ra$ number we do not find any evidence for a sudden onset around this point, see Fig. \ref{Fig_SCL09_model} where the vertical lines are plotted at the same positions as in Fig. \ref{Fig_SCL09_flow}. In contrast to the decrease in the volume averaged value of $Re_{z,rms}$ the horizontally averaged value of $Re_{z,rms}$ at the edge of the thermal BL (thus at the distance $\lambda^{sl}_{\theta}(r)$ from either the top or bottom boundary) increases. This indicates that Ekman pumping, which is responsible for the increase in $Nu$, sets in and no sign of Ekman pumping prior to the decrease in LSC strength is found. Fig. \ref{Fig_SCL09_flow} thus shows that the flow makes a transition between two different turbulent states, i.e. a transition from a LSC dominated regime to an Ekman pumping dominated regime \cite{ste09}. Fig. \ref{Fig_SCL09_flow}a shows no increase in the horizontally averaged value of $Re_{z,rms}$ at the edge of the thermal BL for $Pr=0.7$, because the flow is suppressed for higher $Ro$, i.e. lower rotation rate, when $Pr$ is lower, see the discussion in Ref. \cite{ste09d}.

The $Pr$ number dependence of the $Nu$ number and the thickness of the thermal BL is shown in Fig. \ref{Fig_SCL09_model_Pr}. From Fig. \ref{Fig_SCL09_model_Pr}a we can conclude that hardly any $Pr$ number dependence on the $Nu$ number exists in the weak rotating regime ($Ro=3$). However, a strong $Pr$ number effect appears for stronger rotation rates, where Ekman pumping is the dominant effect \cite{zho09b}. Fig. \ref{Fig_SCL09_model_Pr}b shows that the effect of weak background rotation on the thermal BL thickness is largest for $Pr \approx 2$. The $Pr$ number dependence of $Re_{z,rms}$ is shown in Fig.\ \ref{Fig_SCL09_flow_Pr}. The difference between the data points obtained for the bottom and top BL in Fig.\ \ref{Fig_SCL09_flow_Pr} indicate the uncertainty in the results. Increasing the averaging time, which we checked for $Pr=2$ and $Pr=4.4$, reduces the differences for the data points obtained for the bottom and top BL. Furthermore, we note that it is important to take the radial thermal BL dependence ($\lambda_\theta^{sl}(r)$) into account for lower $Pr$, where the radial BL dependence is strongest, and we excluded the region close to the sidewall ($0.45<r<0.5$) from the horizontal averaging in order to eliminate the effect of the sidewall.

\begin{figure}
  \centering
  \subfigure[]{\includegraphics[width=0.48\textwidth]{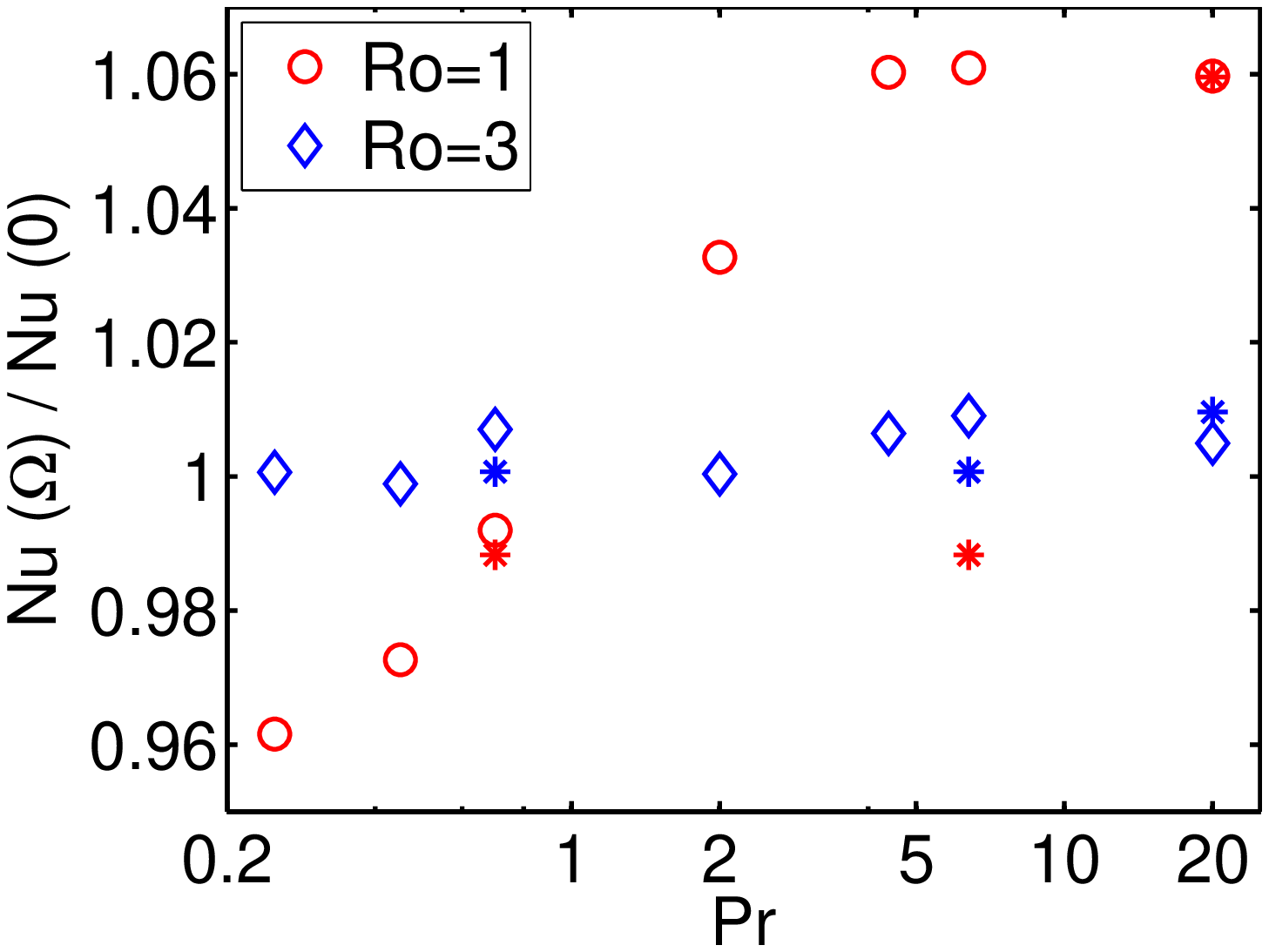}}
  \subfigure[]{\includegraphics[width=0.48\textwidth]{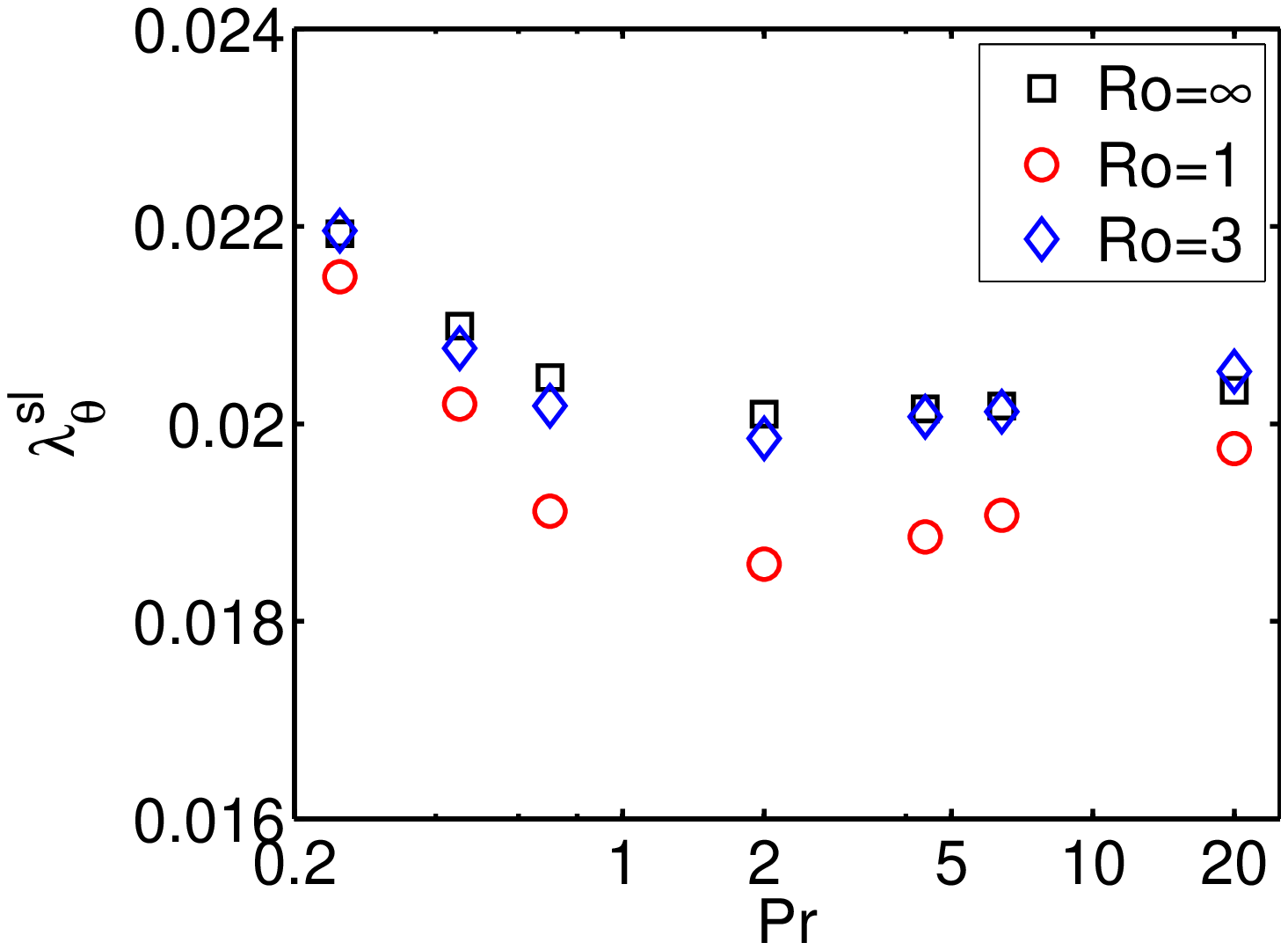}}
  \caption{a) Normalized heat transfer as a function of $Pr$ number for $Ra=4\times10^7$ and different $Ro$. Red open circles and blue diamonds indicate the data for $Ro=1$, and $Ro=3$, respectively. The symbols (circles and diamonds) and stars indicate the results obtained on the $97 \times 257 \times 193$ and the $65 \times 193 \times 129$ grid, respectively. b) Horizontally averaged thermal BL thickness $\lambda^{sl}_{\theta}$ as function of $Pr$. Symbols as in a) and black squares for $Ro=\infty$.}
  \label{Fig_SCL09_model_Pr}
\end{figure}

\begin{figure}
  \centering
  \subfigure[]{\includegraphics[width=0.40\textwidth]{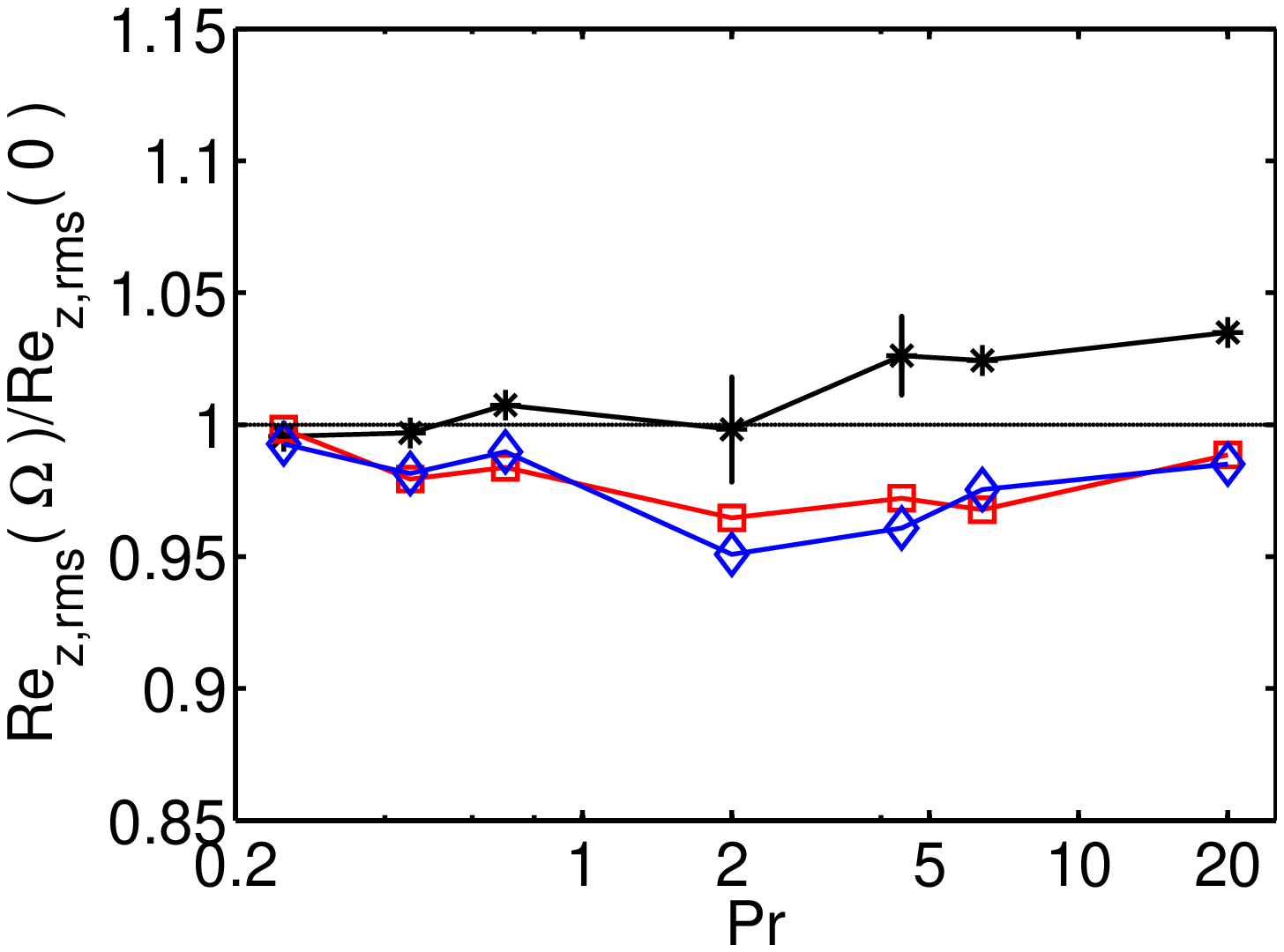}}
  \subfigure[]{\includegraphics[width=0.40\textwidth]{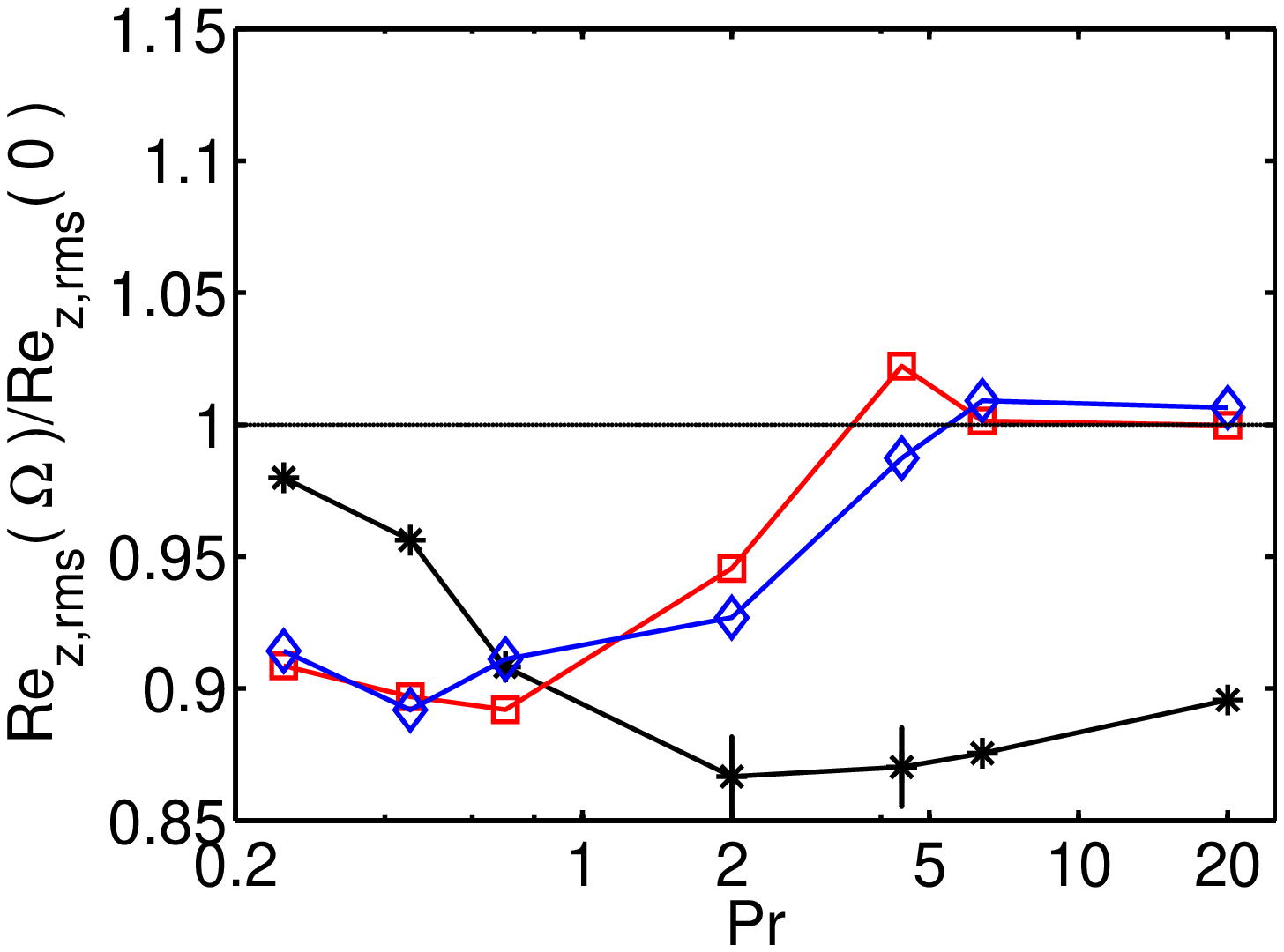}}
  \caption{The normalized averaged rms vertical velocity fluctuations $Re_{z,rms}$ for $Ra=4\times 10^7$ and different $Ro$ as function of $Pr$. The black line indicates the volume averaged value of $Re_{z,rms}$. The vertical error bars at $Pr=2$ and $Pr=4.4$ indicate the difference in the volume averaged value obtained after $1200$ dimensionless time units (data point) and $400$ dimensionless time units. The red and the blue line indicate the horizontally averaged $Re_{z,rms}$ at a distance $\lambda_{\theta}^{sl}(r)$ from the lower and upper plate, respectively. a) $Ro=3$, b) $Ro=1$.}
  \label{Fig_SCL09_flow_Pr}
\end{figure}

\begin{figure}
  \centering
  \subfigure[]{\includegraphics[width=0.32\textwidth]{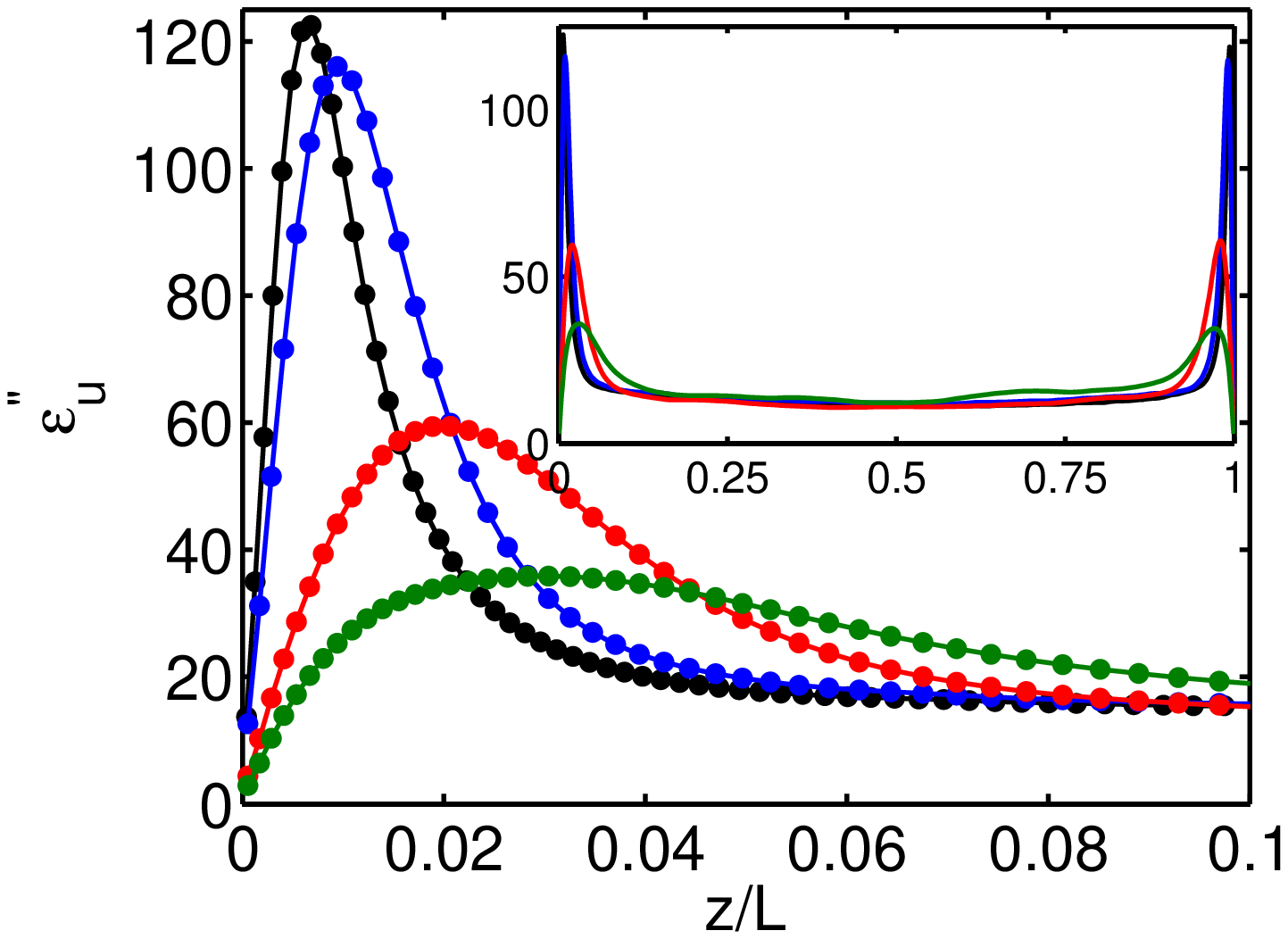}}
  \subfigure[]{\includegraphics[width=0.32\textwidth]{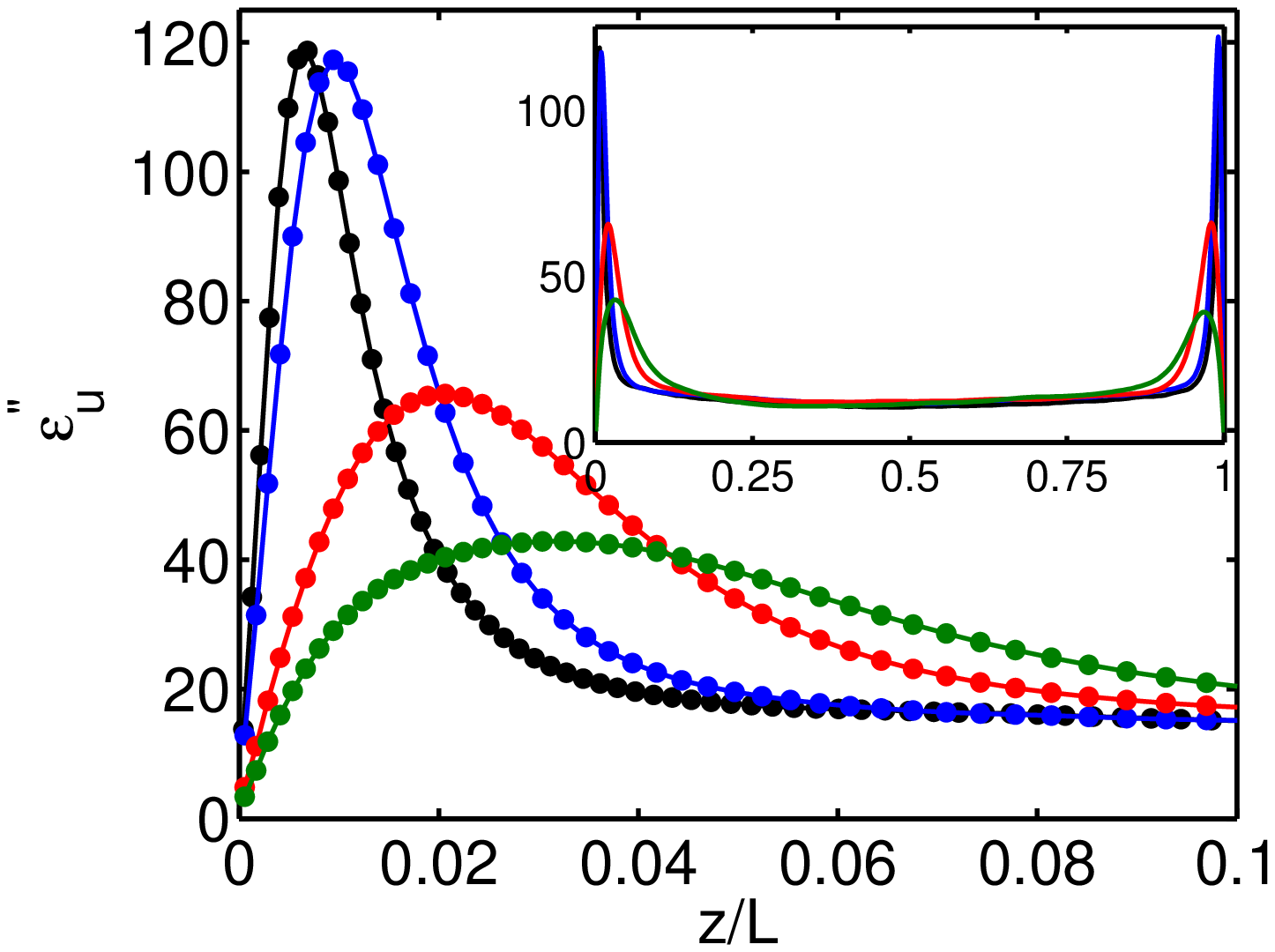}}
  \subfigure[]{\includegraphics[width=0.32\textwidth]{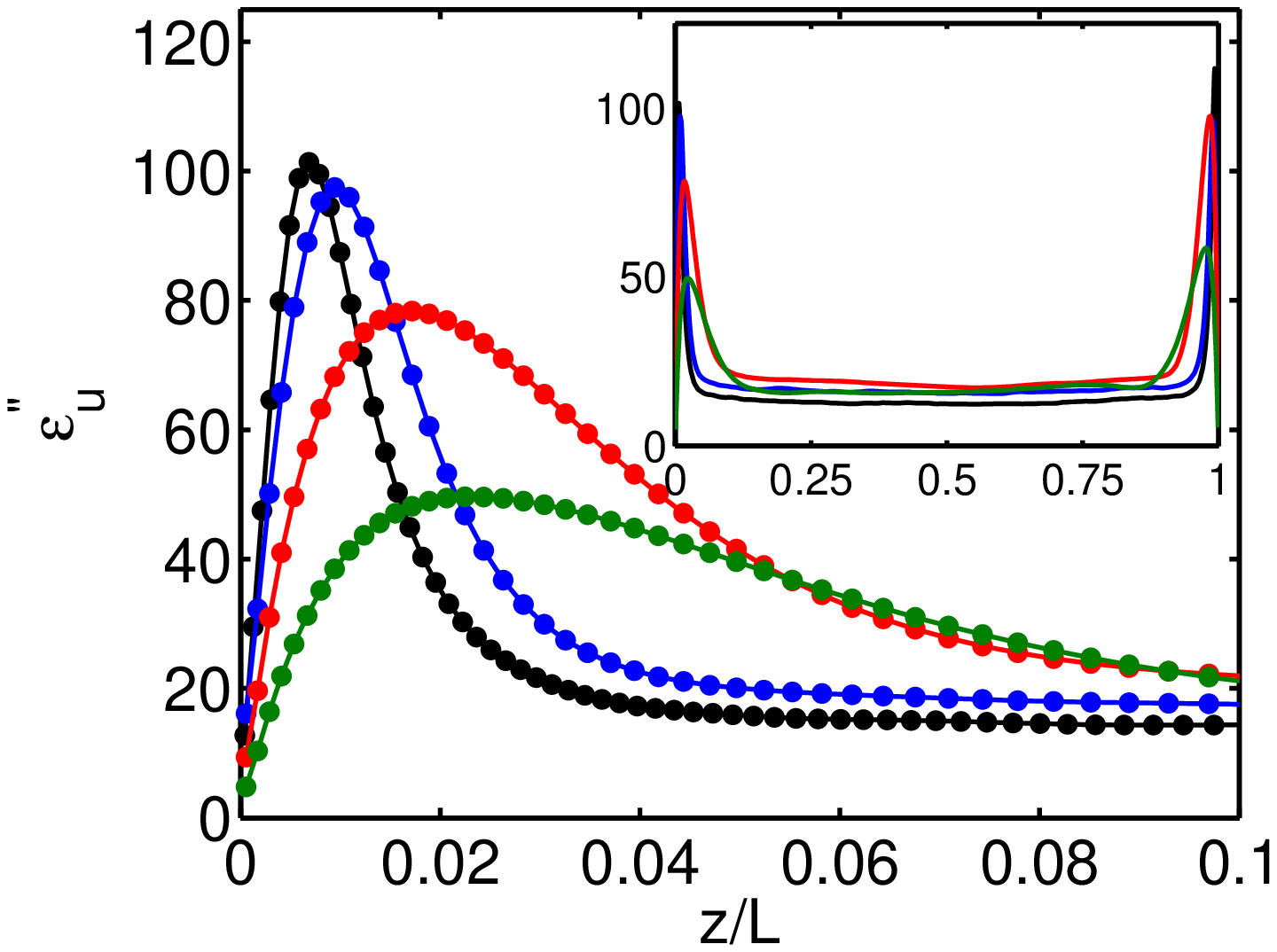}}
  \caption{Azimuthally averaged profile of $\epsilon_u^"$ for $Ra=4\times10^7$ and different $Pr$ and $Ro$ at the radial position $r=0.25L$. The figures, from left to right, are for $Ro=\infty$, $Ro=3$, and $Ro=1$. Black, blue, red, and green indicate the profiles for $Pr=0.25$, $Pr=0.7$, $Pr=6.4$, and $Pr=20$, respectively. The dots indicate the data points obtained from the simulations. The insets shows the profile over the whole domain.}
  \label{Fig_SCL09_profile_epsilonu}
\end{figure}

\begin{figure}
  \centering
  \subfigure[]{\includegraphics[width=0.48\textwidth]{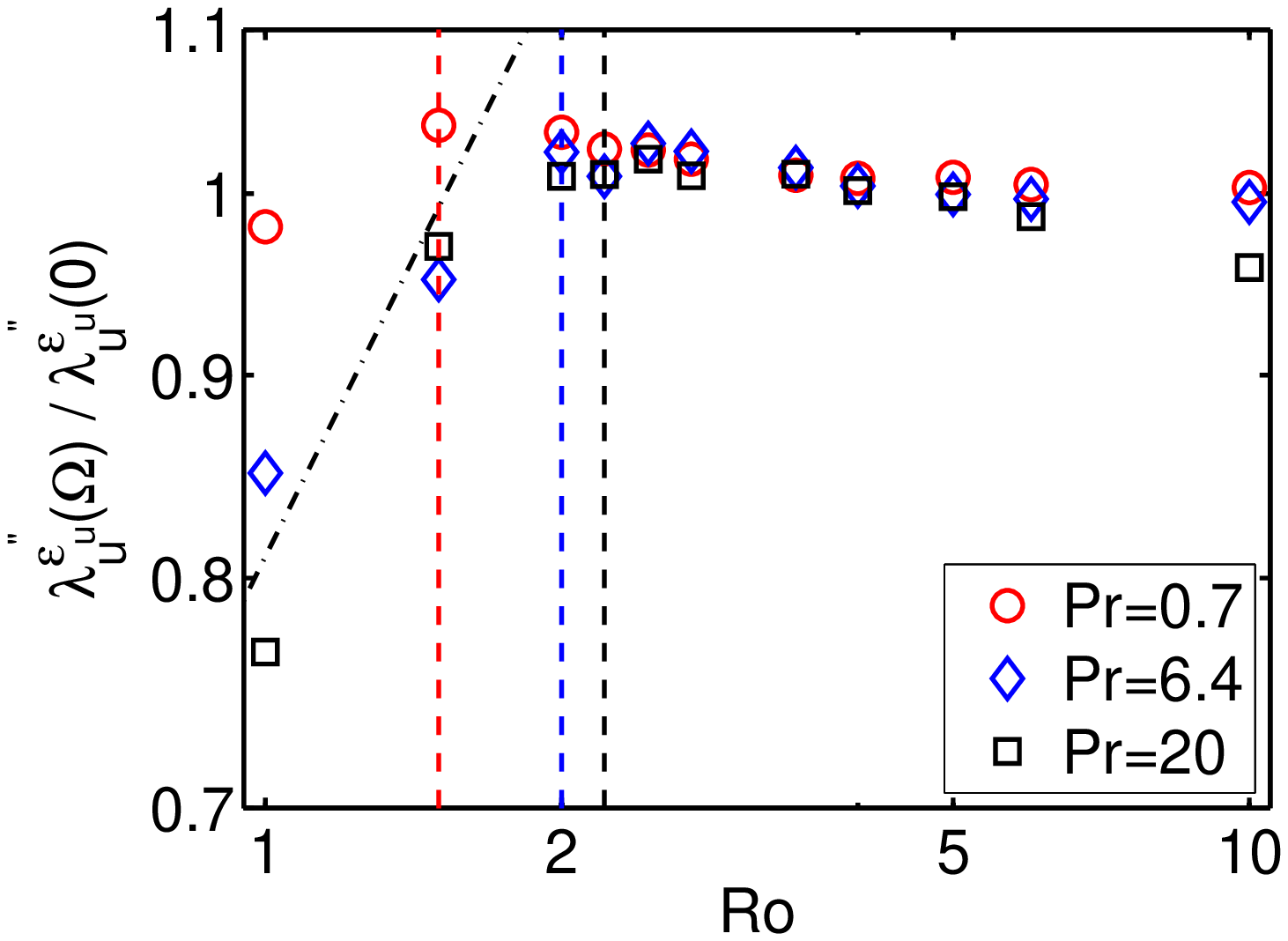}}
  \subfigure[]{\includegraphics[width=0.48\textwidth]{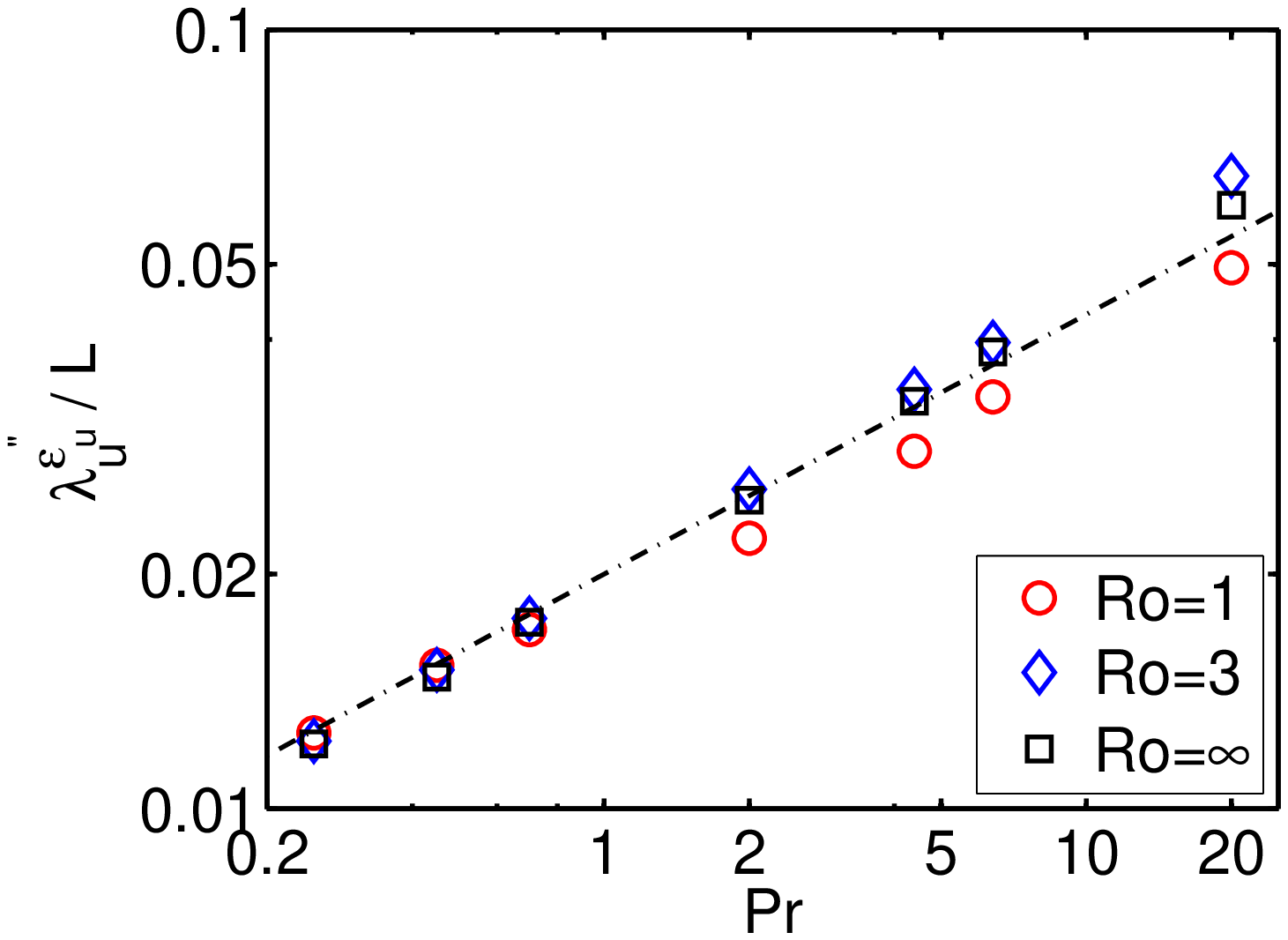}}
  \caption{a) The thickness of the kinetic BL, average of bottom and top BL, based on the position of the maximum value of $\epsilon_u^"$ for $Ra=4\times10^7$ and different $Pr$. Red open circles, blue diamonds, and black squares are the data for $Pr=0.7$, $Pr=6.4$, and $Pr=20$, respectively.  The dash-dotted line indicates $Ro^{1/2}$ scaling. b) As in a), but now for different $Ro$. Red open circles, blue diamonds, and black squares are the data for $Ro=1$, $Ro=3$, and $Ro=\infty$, respectively. The dash-dotted line indicates $Pr^{1/3}$ scaling.
}
  \label{Fig_SCL09_kineticBL}
\end{figure}

The computation of the kinetic BL thickness can be either based on the position of the maximum rms value of the azimuthal velocity fluctuations \cite{ste09,kun10}, or on the position of the maximum value of $\epsilon_u^":= \bf{u} \cdot \nabla^2 \bf{u}$, i.e. two times the height at which this quantity is highest, as shown in \cite{ste09b}. In ref. \cite{lak10} we will compare in detail the profiles of $\epsilon_u$ and $\epsilon_u^"$ and will show that the latter is indeed suited to define the BL thickness. Here we first average $\epsilon_u^"$ horizontally in the range $0.05\le r \le 0.45$ before we determine the position of the maximum. This $r$ range has been taken to exclude the region close to the sidewall, where $\epsilon_u^"$  misrepresents the kinetic BL thickness due to the rising plumes, and the region close to the cylinder axis, since there it is numerically very difficult to reliably calculate $\epsilon_u^"$, due to the singularity in the coordinate system. When the radially dependent kinetic BL thickness ($\lambda_u^{\epsilon_u^"}(r)$) is horizontally averaged a small difference, depending on the averaging time, is observed between the bottom and top because of the specific orientation of the LSC.  We note that the same quantity $\epsilon_u^"$ is used in Ref. \cite{ste09b}, where it is shown that the kinetic BL thickness based on $\epsilon_u^"$ represents the BL thickness better than the one considering the maximum rms velocity fluctuations, which is normally used in the literature. The volume averaged value of $\epsilon_u^"$ is the same as the volume averaged kinetic energy dissipation rate $\epsilon_u$ (although it differs locally), which can easily be derived using Gauss's theorem \cite{lak10}. Fig. \ref{Fig_SCL09_profile_epsilonu} shows the azimuthally averaged profiles for $\epsilon_u^"$ at $r=0.25L$ and in Fig. \ref{Fig_SCL09_kineticBL} the kinetic BL thickness based on the position of the maximum kinetic dissipation rate is shown as function of $Ro$ and $Pr$. To compare the relative changes in the kinetic BL thicknesses the values are normalized by values for the non-rotating case. For all $Pr$ numbers there is a change in the BL behaviour at the point where the LSC decreases in strength (vertical dashed lines in Fig. \ref{Fig_SCL09_kineticBL}a).

\begin{figure}[!hbt]
  \centering
  \subfigure[]{\includegraphics[width=0.32\textwidth]{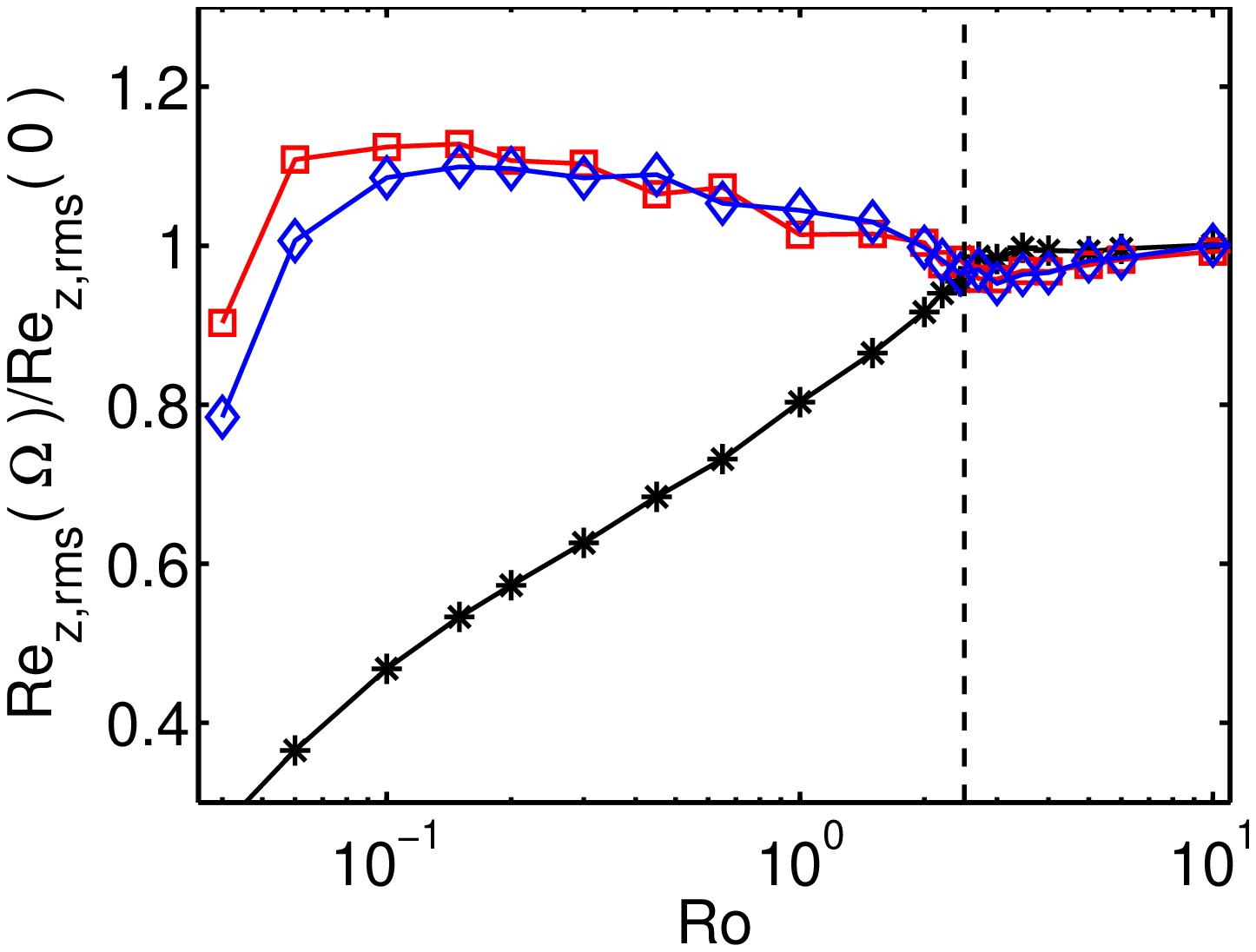}}  \subfigure[]{\includegraphics[width=0.32\textwidth]{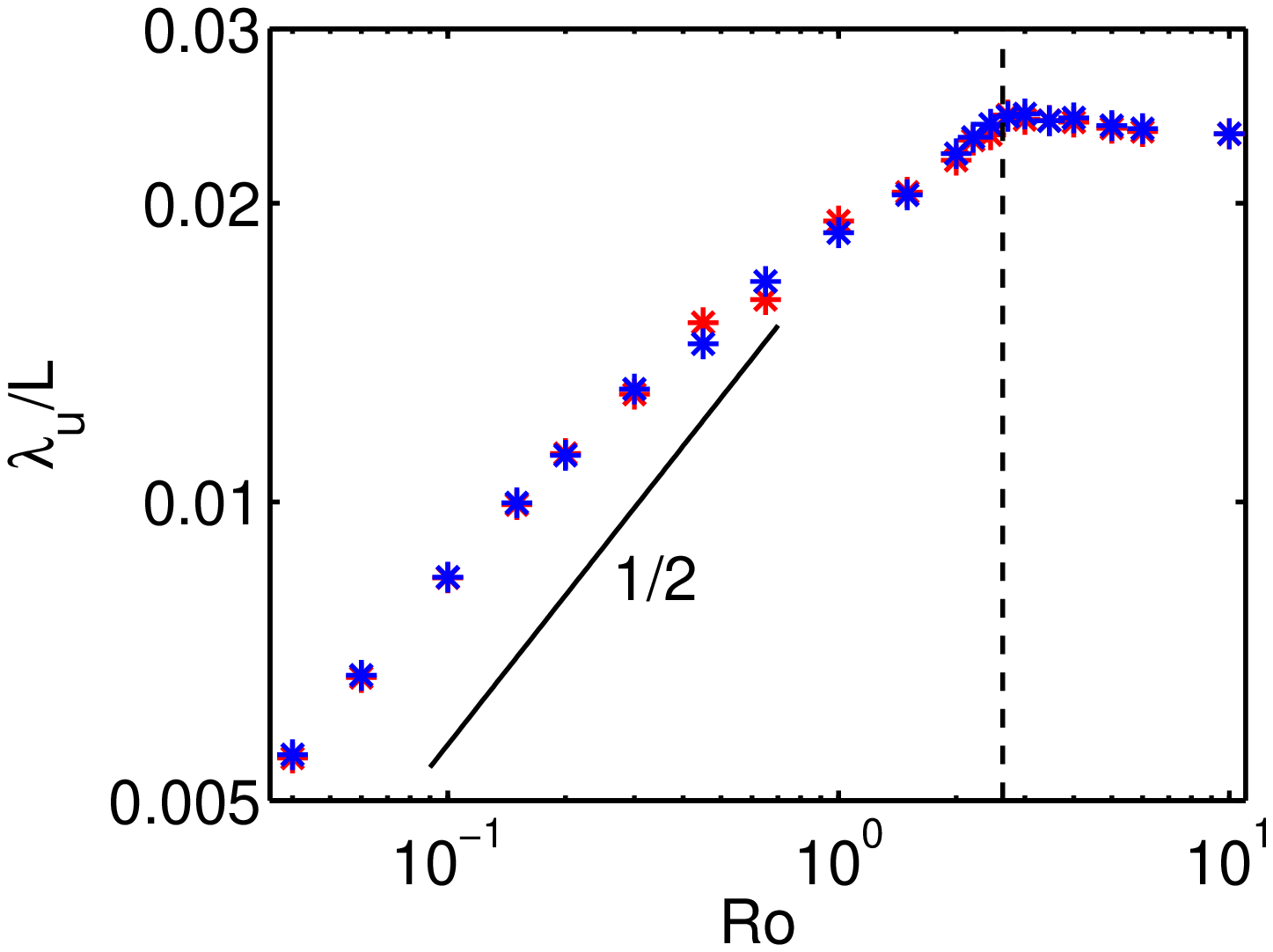}}
  \subfigure[]{\includegraphics[width=0.32\textwidth]{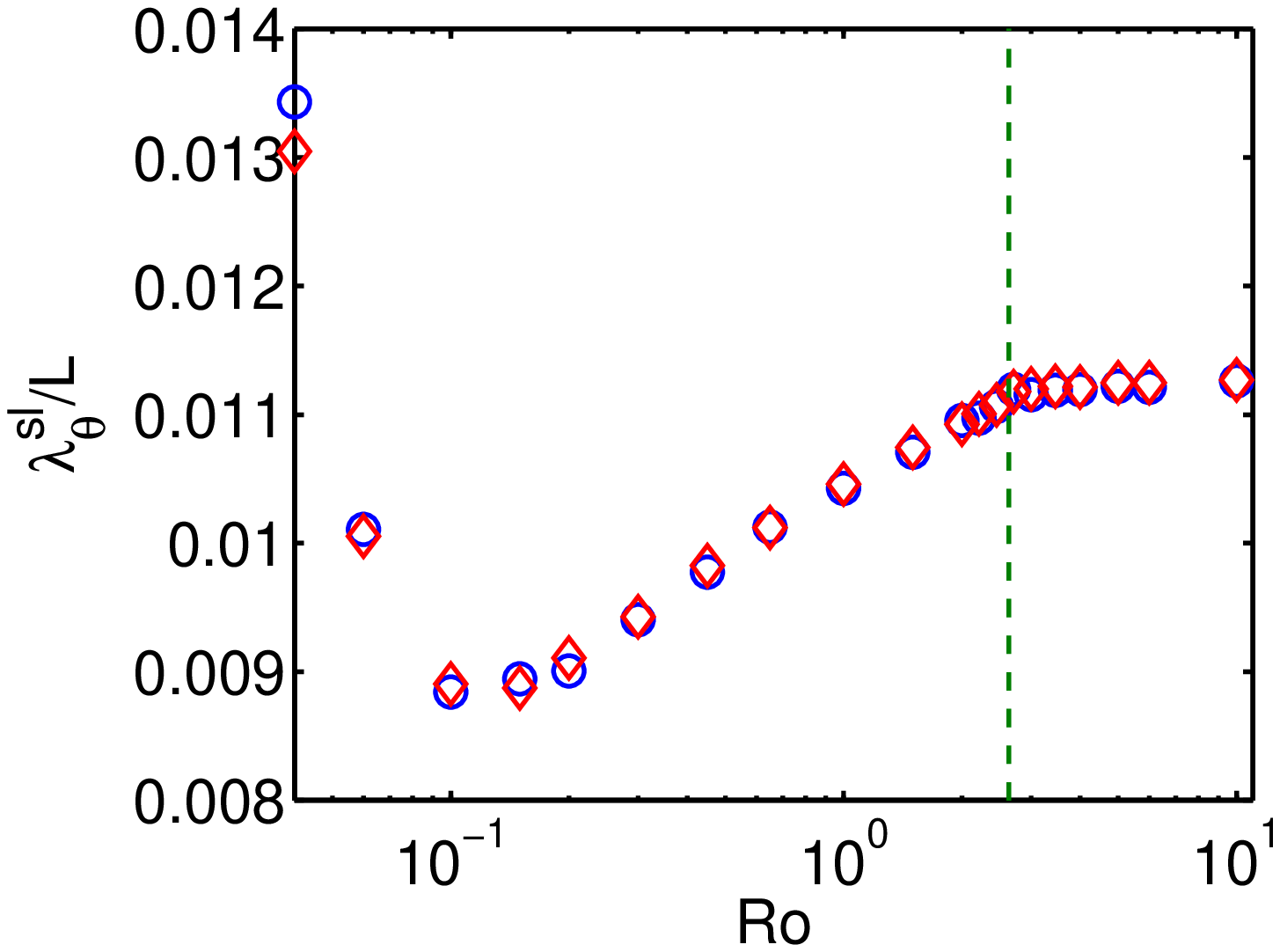}}
  \caption{a) The averaged rms vertical velocities $Re_{z,rms}$ as function of $Ro$. The black line indicates the volume averaged value of $Re_{z,rms}$. The red and the blue line, respectively, indicate the horizontally averaged $Re_{z,rms}$ at the distance $\lambda_\theta^{sl}(r)$ from the bottom and top plate. b) The thickness of the kinetic top and bottom BLs based on the position of the maximum $\epsilon_u^"$, blue (red) for top (bottom) BL. The vertical dashed lines in panel a and b indicate the position of  $Ro_c$. Here the BL behaviour changes from Prandtl-Blasius (right) to Ekman (left) \cite{ste09}. c) The blue circles (red diamonds) indicate $\lambda_\theta^{sl}$ for the bottom (top) plate.}
  \label{Fig_SCL09_highRa2}
\end{figure}

We conclude this section with a brief summary of the results obtained for the high $Ra$ number regime \cite{zho09b,ste09}. For $Ra\gtrsim 1\times10^8$ a sudden onset at $Ro=Ro_c$ in the heat transport enhancement occurs, see Fig. \ref{Fig_SCL09_highRa_a}b, where we find a smooth transition at lower $Ra$, see Fig. \ref{Fig_SCL09_highRa_a}a. Fig. \ref{Fig_SCL09_highRa2}a shows the volume averaged ratio $Re_{z,rms}(\Omega)$/$Re_{z,rms}(0)$. The behaviour is similar to the one observed at lower $Ra$, see Fig. \ref{Fig_SCL09_flow}. For the high $Ra$ number regime the onset at $Ro_c$ is defined as the point where the ratio $Re_{z,rms}(\Omega)$/$Re_{z,rms}(0)$ becomes smaller than $1$. For the low $Ra$ number this point indicates the position where the LSC strength starts to decrease and just as for the relatively low $Ra$ number regime, Ekman pumping in the high $Ra$ number regime is indicated by an increase of the horizontally averaged $Re_{z,rms}$ value at the edge of the thermal BL. Although the two cases, i.e. the relatively low $Ra$ number regime and the high $Ra$ number regime both, show a transition between two different turbulent states the important difference between the two is that the transition is sharp in the high $Ra$ number regime and smooth in the relatively low $Ra$ number regime.
The onset in the high $Ra$ number regime is also observed in the behaviour of the BLs. Fig. \ref{Fig_SCL09_highRa2}b shows that the kinetic BL thickness does not change below onset ($Ro> Ro_c$) and above onset the BL behaviour is dominated by rotational effects and thus Ekman scaling (proportional to $Ro^{1/2}$) is observed. This scaling factor is also found in the laminar BL theory, which will be discussed in the next section. Finally, fig. \ref{Fig_SCL09_highRa2}c shows the thermal BL thickness $\lambda_\theta^{sl}$.

\section{Boundary layer theory for weak background rotation}
\label{Sec3}

In the previous section we observed that there is a smooth increase in the heat transfer as function of the $Ro$ number when the $Ra$ number is relatively low. In this section we set out to account for the increase in the heat transfer as function of the rotation rate within a model, which extends the ideas of the GL-theory to the rotating case. In the GL theory the Prandtl-Blasius BL theory for laminar flow over an infinitely large plate was employed in order to estimate the thicknesses of the kinetic and thermal BLs, and the kinetic and thermal dissipation rates. These results were then connected with the $Ra$ and $Pr$ number dependence of the Nusselt number. In perfect analogy, in the present paper we apply laminar BL theory for the flow over an infinitely large rotating plate to study the effect of rotation on the scaling laws. We stress that employing the results of laminar BL theory over an infinite rotating disk to the rotating RB case in a closed cylinder is fully analogous to employing Prandtl-Blasius BL theory for flow over an infinite plate to the standard RB case without rotation, where the method was very successful \cite{gro00,gro01,gro02,gro04,ahl09}.

In both cases the equations used to derive the scaling laws are time independent and therefore the resulting solutions are associated with laminar flow. However, evidently, high Rayleigh number thermal convection is time dependent. Therefore one wonders whether the derived scaling laws still hold for time-dependent flow over an infinite rotating disk. We will show that the $Ro$ and $Pr$ scaling that is derived is not changed when temporal changes are included. This is again in perfect analogy to the Prandtl-Blasius BL case where the scaling laws also hold for time dependent flow provided that the viscous BL does not break down \cite{gro04}. Indeed, recent experiments and numerical simulations \cite{sun08,zho09c,zho10} have shown that in non-rotating RB the BLs scaling wise behave as in laminar flow and therefore we feel confident to assume the same for the weakly rotating case. The basic idea of the model we introduce is to combine the effect of the LSC roll, which is implemented in the GL theory by the use of laminar Prandtl-Blasius BL theory over an infinitely large plate, and the influence of the rotation on the thermal BL.

\begin{figure}
   \centering
   \subfigure[]{\includegraphics[width=0.48\textwidth]{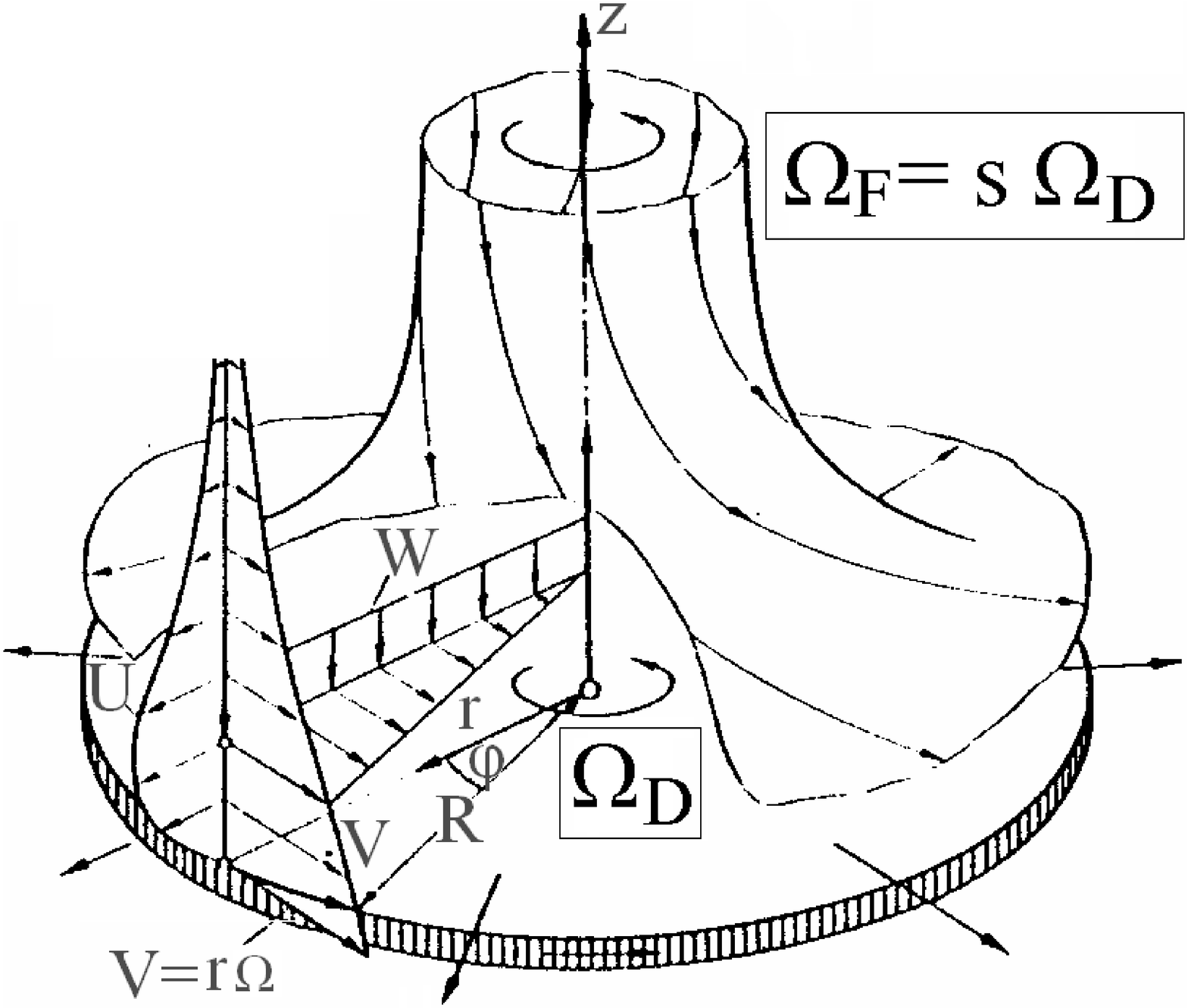}}
   \subfigure[]{\includegraphics[width=0.48\textwidth]{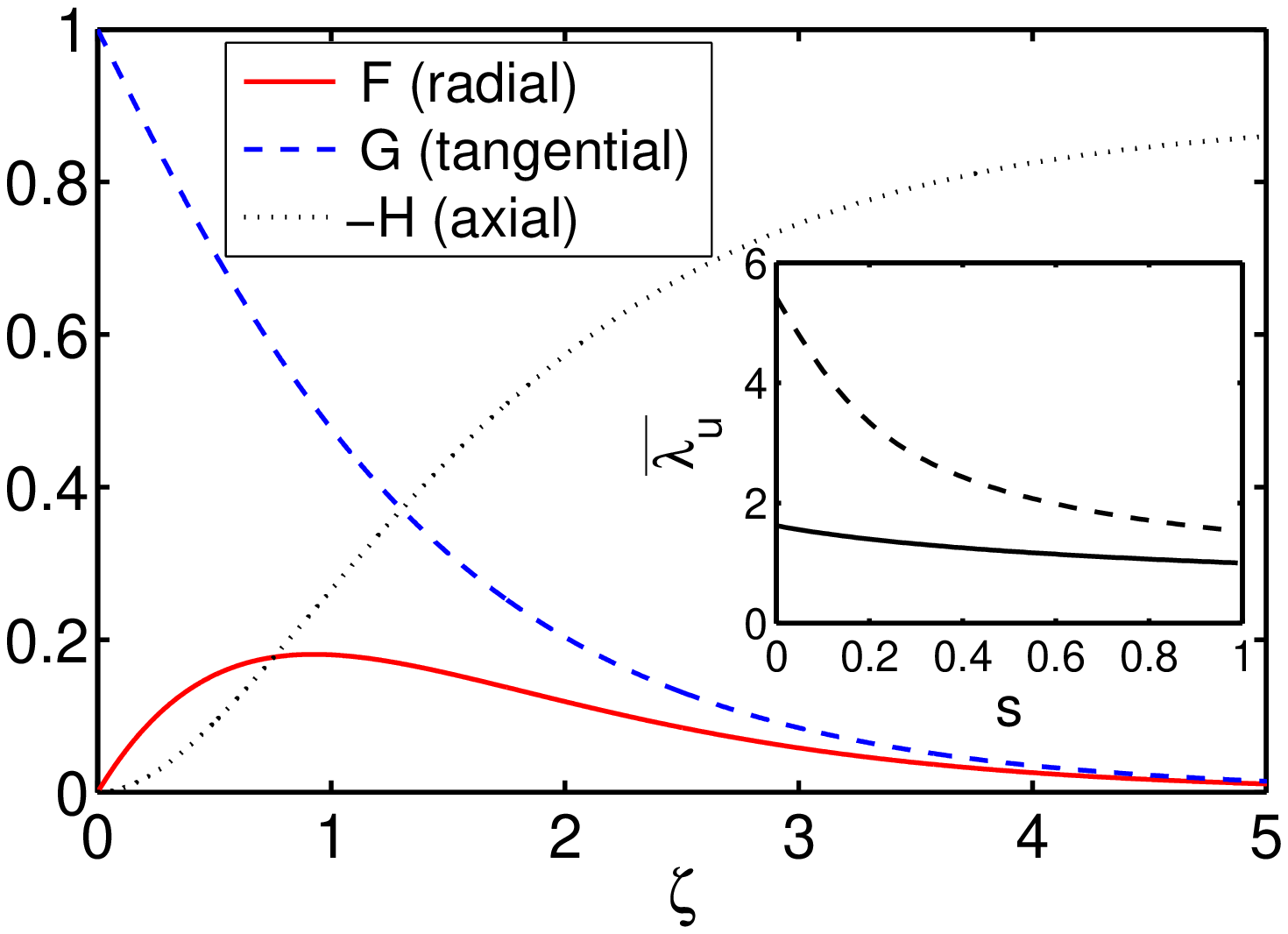}}
   \caption{a) Flow near a disk rotating with an angular velocity $\Omega_D$ when the fluid at infinity is rotating with $\Omega_F = s \Omega_D$. Adapted from \cite{sch79}. b) Velocity components in the von K\'arm\'an case: inset shows $\overline{\lambda}_u^{sl}$ (solid line) and $\overline{\lambda}_u^{99\%}$ (dashed line).}
   \label{Figure_ste08_setup}
 \end{figure}

The system we are analyzing to study the influence of rotation on the thermal BL thickness above a heated plate is schematically shown in Fig. \ref{Figure_ste08_setup}a. It is the laminar flow of fluid over an infinite rotating disk. The disk rotates with an angular velocity $\Omega_D$ and the fluid at infinity with angular velocity $\Omega_F=s\Omega_D$, with $s<1$. Fig. \ref{Figure_ste08_setup}b shows that a positive radial velocity is created due to the action of the centrifugal force. Because of continuity there is a negative axial velocity, i.e. fluid is flowing towards the disk.
The system has been analyzed before in the literature, e.g. Refs. \cite{loi67,rog60,sch79,wij85,zan77}. Here we will briefly summarize the procedure.

The system is analyzed by using the Navier-Stokes equations in cylindrical coordinates and assuming a steady stationary, axial symmetric solution. To reduce the Navier-Stokes equations to a set of ordinary differential equations (ODE) we employ self-similarity. The first step is to determine the dimensionless height in the system just as in the Prandtl-Blasius approach, in which the BL thickness scales as \cite{ll87} $\delta \sim \sqrt{\nu x/U}$. For the case of a large rotating disk in a fluid rotating around an axis perpendicular to the disk, the thickness of the BL can be estimated by replacing $U$ by $\Omega_D x$ \cite{ll87}. The thickness then scales as $\delta \sim \sqrt{\nu/\Omega_D}$. The similarity variable in the system is the dimensionless height

\begin{equation}\label{Eq dimensionless height}
    \zeta=z \sqrt{\frac{\Omega_D}{\nu}}.
\end{equation}
According to von K\'arm\'an, the following self-similarity ansatz for the velocity components and the pressure
 can be taken \cite{loi67,rog60,sch79,wij85,zan77}
\begin{eqnarray}
    \label{Eq assumption radial velocity}
    u &=& r \Omega_D F(\zeta),\\
    \label{Eq assumption tangential velocity}
    v &=& r \Omega_D G(\zeta),\\
    \label{Eq assumption axial velocity}
    w &=& \sqrt{\nu \Omega_D} H(\zeta),\\
    \label{Eq assumption pressure}
    p &=& \rho \nu \Omega_D P(\zeta) + \frac{1}{2} \rho s^2 \Omega_D^2 r^2.
\end{eqnarray}
After substitution into the Navier-Stokes equations one obtains a system of four coupled ODEs,
\begin{eqnarray}
    \label{Eq radial momentum}
    &&F^2 + F^{'}H - G^{2} -F^{''} + s^2 = 0, \\
    \label{Eq tangential momentum}
    &&2FG + HG^{'} - G^{''} = 0,  \\
    \label{Eq pressure}
    &&P^{'}+HH^{'}-H^{''} = 0,\\
    \label{Eq continuity}
    &&2F + H^{'} = 0,
\end{eqnarray}
where the prime indicates differentiation with respect to $\zeta$. This set of ODEs must be supplemented by the boundary conditions
\begin{eqnarray}
    \label{Eq boundary condition dimensional z0}
    u = 0,\hspace{3mm}  v=  r \Omega_D &,\hspace{3mm}  w=0 \hspace{3mm}  &\mbox{ for } z=0,\\
    \label{Eq boundary condition dimensional zinfty}
    u = 0,\hspace{3mm}  v=s r \Omega_D &\hspace{3mm}                    &\mbox{ for } z=\infty.
\end{eqnarray}
When substituting the self similarity ansatz (\ref{Eq assumption radial velocity}) - (\ref{Eq assumption pressure}) into these boundary conditions one obtains
\begin{eqnarray}
    \label{Eq boundary condition nondimensional z0}
    F = 0,\hspace{3mm}  G=1 &,\hspace{3mm}    H=0 \hspace{3mm} & \mbox{ for } \zeta=0,\\
    \label{Eq boundary condition nondimensional zinfty}
    F = 0,\hspace{3mm}  G=s &\hspace{3mm}         & \mbox{ for } \zeta=\infty.
\end{eqnarray}
Note that the boundary condition at infinity together with the continuity equation (\ref{Eq continuity}) gives $H^{'}(\zeta\rightarrow\infty)=0$. One can further simplify the set of ODEs by realizing that the ODE for the pressure (\ref{Eq pressure}) is decoupled from the ODEs determining the velocity profiles by using the continuity equation (\ref{Eq continuity}) in (\ref{Eq pressure}) and subsequently integrating this relation. The velocity profiles, for the von K\'arm\'an case ($s=0$), are shown in Fig. \ref{Figure_ste08_setup}b. The inset shows that the dimensionless kinetic BL thickness $\overline{\lambda}_u \equiv {\lambda_u}/\delta$ decreases with increasing relative rotation rate $s$ of the fluid at infinity. This is due to the decreasing effect of the centrifugal force. We determined $\overline{\lambda}_u^{99\%}$, the dimensionless kinetic BL thickness at which the velocity has achieved $99\%$ of the outer flow velocity, using the tangential velocity profile, i.e. when $G(\zeta)=s+0.01(1-s)$. Additionally, we calculated $\overline{\lambda}_u^{sl}$, the dimensionless kinetic BL thickness based on the slope of the tangential velocity at the disk. The scaling of the kinetic BL predicted by the above rotating BL theory, i.e. $Ro^{1/2}$, is the classical Ekman scaling. In the simulations of RRB we find the same scaling of the kinetic BL once the flow is dominated by rotational effect, i.e. $Ro \lesssim Ro_c$, see Fig. \ref{Fig_SCL09_highRa2}b.

The GL theory heavily builds on laminar Prandtl-Blasius BL theory, which describes the laminar flow over an infinite plate. In the Prandtl-Blasius theory the temperature field is assumed to be passive to derive the $Pr$ number scaling. As we want to extend the GL theory to the rotating case we keep this analysis analogous to the Prandtl-Blasius theory. Therefore, we assume the temperature field to be passive in order to derive the scaling laws as function of the $Pr$ number. We non-dimensionalize the temperature by

\begin{equation}\label{Eq definition dimensionless temperature}
    \widetilde{\theta}(\zeta)= \frac{\theta-\theta_\infty}{\theta_{b}-\theta_{\infty}},
\end{equation}
where $\theta_b$ is the temperature of the bottom disk, and $\theta_\infty<\theta_b$ is the ambient temperature. Then one obtains the following ODE describing the temperature field \cite{spa59,vir80,mil52}
\begin{equation}\label{Eq ODE energy equation}
    \widetilde{\theta}^{''} = Pr H(\zeta) \widetilde{\theta}',
\end{equation}
where the prime indicates a differentiation with respect to $\zeta$. The boundary conditions are
\begin{eqnarray}
    \label{Eq boundary condition thermal z0}
    \widetilde{\theta} = 1 &  \mbox{ for } & \zeta = 0,\\
    \label{Eq boundary condition thermal zinfty}
    \widetilde{\theta} = 0 &  \mbox{ for } & \zeta = \infty.
\end{eqnarray}
The resulting system of ODEs subjected to the boundary conditions, is solved numerically with a fourth order Runge-Kutta method using a Newton-Raphson root finding method to find the initial conditions. One can take the analytic solution for the Ekman case ($s\approx 1$), see appendix \ref{appA}, or the known solution for the B\"{o}dewadt case, see for example \cite{rog60,sch79}, as one of the starting cases to determine the solutions over the whole parameter range in $s$ and $Pr$.

\begin{figure}
  \centering
  \subfigure[]{\includegraphics[width=0.48\textwidth]{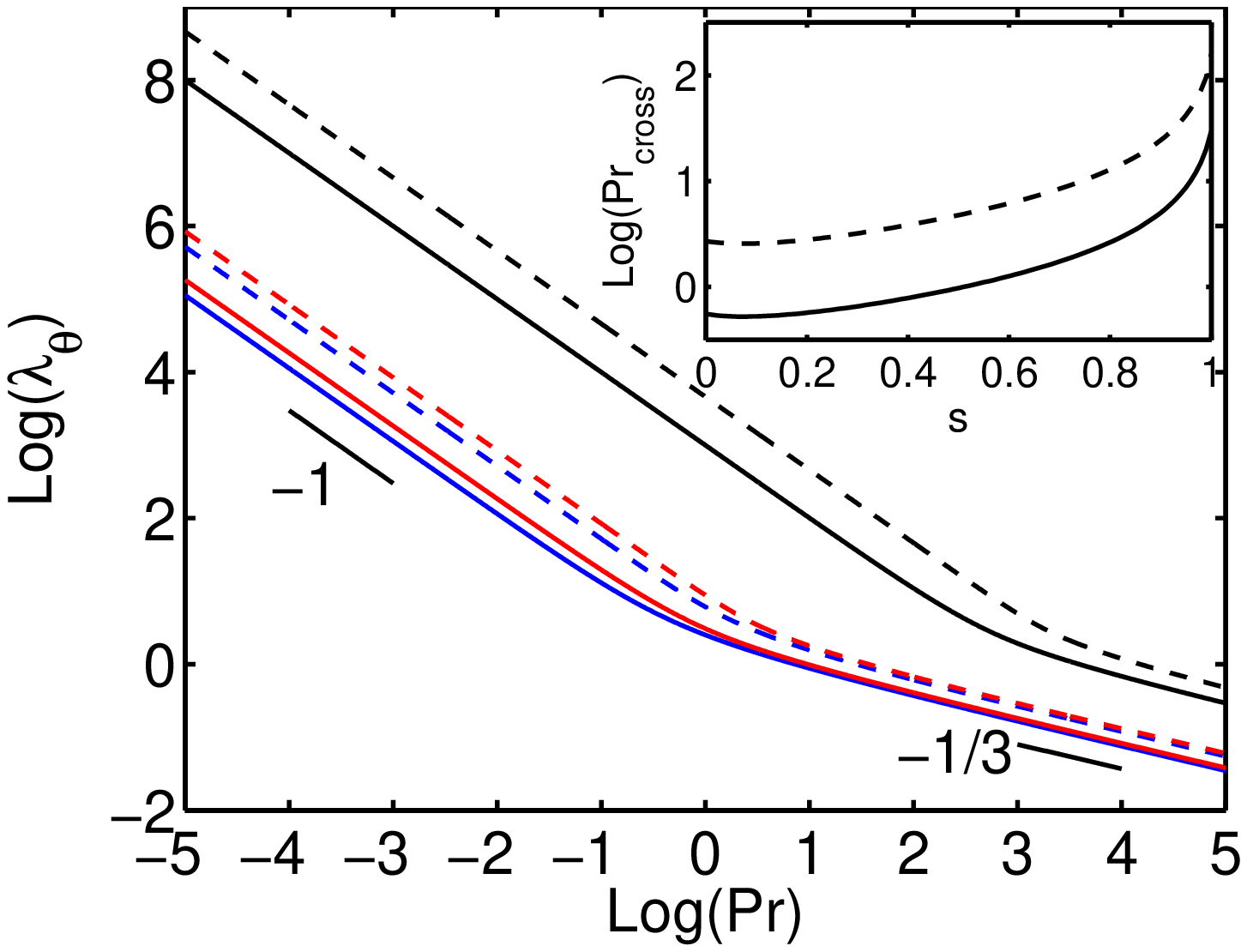}}
  \subfigure[]{\includegraphics[width=0.48\textwidth]{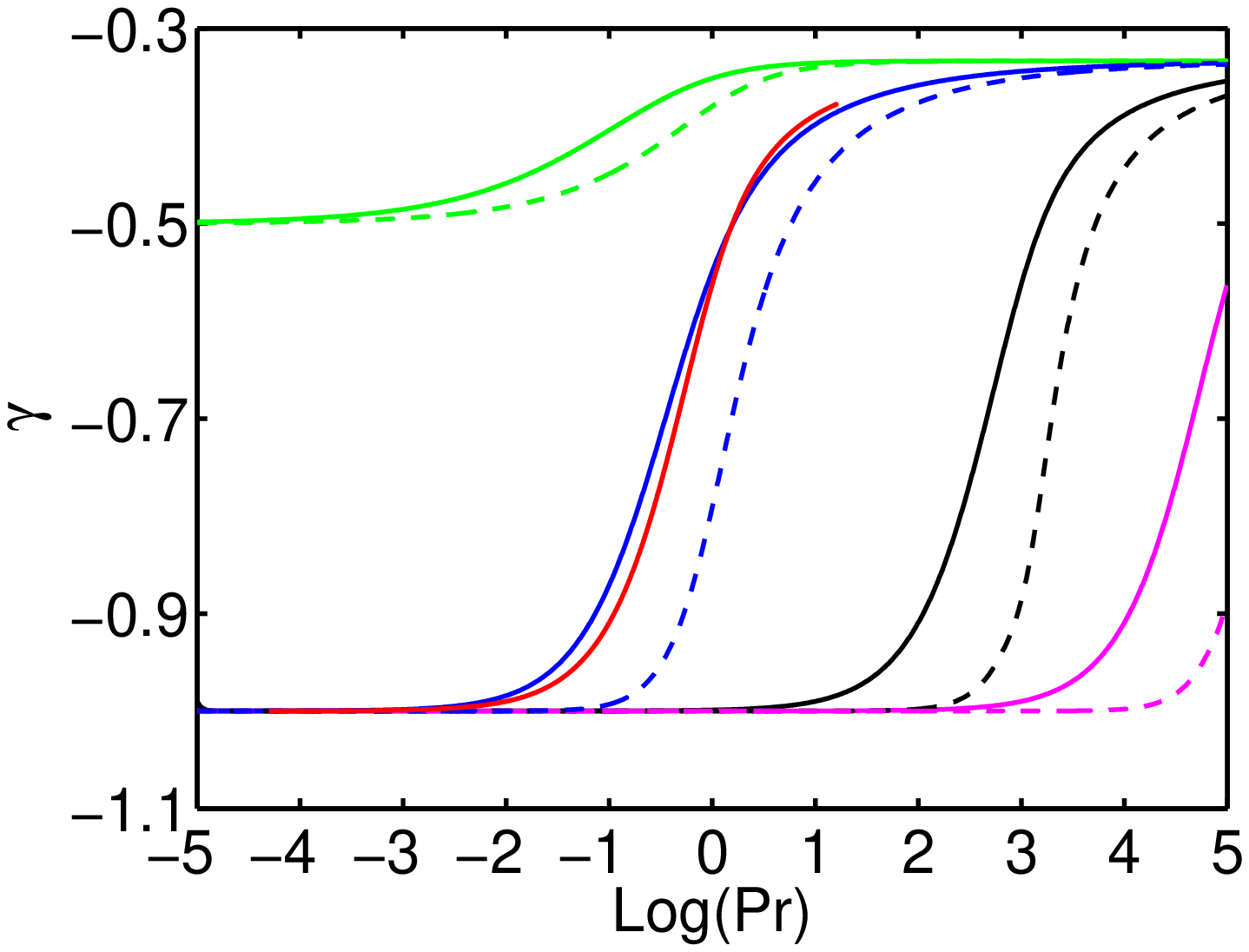}}
  \caption{a) The thermal BL thickness $\overline{\lambda}_\theta \equiv \lambda_\theta/\delta$ as function of $Pr$ for the von K\'arm\'an case ($s=0$) in blue, for $s=0.5$ in red, and for the Ekman case ($s \approx 1, Ro^*= 10^{-3},$ see definition Eq.\ (\ref{Eq Rossbystar})) in black. The solid lines are for $\overline{\lambda}_{\theta}^{sl}$ and the dashed lines for $\overline{\lambda}_{\theta}^{99\%}$. Note that the scaling for the thermal BL thickness goes asymptotically to $Pr^{-1/3}$ in the high $Pr$ regime and to $Pr^{-1}$ in the low $Pr$ regime. The inset shows $Pr_{cross}$, the transition between the high and the low $Pr$ regime as function of $s$ based on the behaviour $\overline{\lambda}_{\theta}^{sl}$ (solid line) and $\lambda_\theta^{99\%}$ (dashed line). Note that the low $Pr$ regime is more favored for higher $s$. b) The effective scaling exponent $\gamma$ in $\overline{\lambda}_\theta \sim Pr^{\gamma}$ as function of $Pr$ for the Blasius case (no rotation, green), the von Karman case ($s=0$, blue), and the Ekman case ($s\approx 1$) for two Rossby numbers  where black indicates the solution for $Ro^*=- 10^{-3}$ and magenta for $Ro^*=- 10^{-5}$. The solid lines indicate the effective scaling for $\overline{\lambda}_{\theta}^{sl}$, and the dotted lines for $\overline{\lambda}_{\theta}^{99\%}$. The red line indicates the analytic prediction, see Eq.\ (\ref{Eq coeff1}), for the von K\'arm\'an case.}
  \label{Figure_ste08_thermalBL}
\end{figure}

In this way we obtain the full temperature profile for all $s$ and $Pr$. For the heat transfer the most relevant quantity is the thermal BL thickness $\lambda_\theta$, which we non-dimensionalized by $\delta$, thus $\overline{\lambda}_\theta \equiv \lambda_\theta/\delta$. One can distinguish between $\overline{\lambda}_{\theta}^{sl}$ 
and $\overline{\lambda}_{\theta}^{99\%}$, the dimensionless BL thickness based on the $99\%$ criterion, thus when $\widetilde{\theta}(\zeta)=0.01$. Fig. \ref{Figure_ste08_thermalBL}a shows that the asymptotic scaling of $\overline{\lambda}_{\theta}^{99\%}$ and $\overline{\lambda}_{\theta}^{sl}$ is the same. Furthermore, the figure shows that rotation does not influence the scaling of the thermal BL thickness in the high $Pr$ regime, because the same scaling, namely proportional to $Pr^{-1/3}$, is found as for the Prandtl-Blasius case. However, the rotation does influence the scaling in the low $Pr$ regime, where now $\overline{\lambda}_{\theta} \propto Pr^{-1}$ instead of $\overline{\lambda}_{\theta} \propto Pr^{-1/2}$ as found in the Prandtl-Blasius case. Notice that $\overline{\lambda}_{\theta}^{99\%} > \overline{\lambda}_{\theta}^{sl}$, which is due to the decreasing temperature gradient with increasing height.

In Fig. \ref{Figure_ste08_thermalBL}b we show the effective power-law exponent $\gamma= (d \log \overline{\lambda}_\theta)/(d \log Pr)$ of an assumed effective power law $\overline{\lambda}_\theta \sim Pr^{\gamma}$. It confirms that the effective scaling in the high $Pr$ regime is the same for the Prandtl-Blasius (no rotation) and the von K\'arm\'an case ($s=0$), but already at $Pr=1$ a significant difference is observed.

The temperature advection equation (\ref{Eq ODE energy equation}) directly suggests the following relation between the scaling of the thermal BL thickness $\overline{\lambda}_{\theta}^{sl}$ and the scaling of the axial velocity at the edge of the thermal BL, $H_{BL}\sim 1/Pr \overline{\lambda}_{\theta}^{sl}$. This immediately implies for the low $Pr$ regime, with $\overline{\lambda}_{\theta}^{sl} \sim Pr^{-1}$, that $H_{BL}$ is independent of $Pr$. For the high $Pr$ regime, with $\overline{\lambda}_{\theta}^{sl} \sim Pr^{-1/3}$, it gives $H_{BL}(Pr) \sim Pr^{-2/3}$. The scaling of the thermal BL thickness in the low $Pr$ regime can be understood on physical grounds. In this regime $\overline{\lambda}_\theta \gg \overline{\lambda}_u$ and, the kinetic BL is fully submerged in the thermal BL. The axial velocity at the edge of the kinetic BL (\ref{Eq boundary condition nondimensional zinfty}) is $H_{BL}=H(\zeta \rightarrow \infty$), as can be shown by applying mass conservation expressed by Eq.\ (\ref{Eq continuity}). As a consequence, in the low $Pr$ regime the axial velocity is constant in almost the whole thermal BL. Then Eq.\ (\ref{Eq ODE energy equation}) can trivially be integrated and immediately gives $\overline{\lambda}_{\theta}^{sl} \sim Pr^{-1}$. This derivation is valid for all $s$, i.e. the scaling in the low $Pr$ regime does not depend on the rotation of the fluid at infinity. The scaling in the high $Pr$ regime is also independent of the rotation of the fluid at infinity. In the Ekman case $s \approx 1$, see appendix \ref{appA}, this $1/3$ scaling regime shifts towards very large $Pr$.

The equations (\ref{Eq radial momentum})-(\ref{Eq continuity}) are time independent and therefore the resulting solutions are understood to describe laminar flow. Temporal changes can easily be included by adding $\partial_{\tilde{t}} \tilde{u}_{\theta}$, $\partial_{\tilde{t}} \tilde{u}_{r}$, and $\partial_{\tilde{t}} \tilde{u}_{z}$ where $\tilde{t}=t\Omega_D$ without changing the $Ro$ and $Pr$ scaling discussed above, i.e. the derived scaling laws above still hold for time dependent flow provided that the viscous BL does not break down. This is in perfect analogy to the Prandtl-Blasius BL case where the scaling laws also hold for time dependent flow \cite{gro04}.

To investigate the crossover between the high and the low $Pr$ regime we define $Pr_{cross}$ as the crossover point. $Pr_{cross}$ is calculated by determining the intersection between the asymptotic behaviour of the high and the low $Pr$ regime. To calculate $Pr_{cross}$ we considered $\overline{\lambda}_{\theta}^{sl}(Pr)$ and $\overline{\lambda}_{\theta}^{99\%}(Pr)$. The inset of Fig. \ref{Figure_ste08_thermalBL}a shows that the low $Pr$ regime becomes more favored when the rotation of the fluid at infinity becomes stronger. Then the kinetic BL thickness becomes thinner and the axial velocity decreases, i.e. the thermal BL thickness increases. Because the thermal BL thickness decreases with increasing $Pr$ the crossover shifts towards higher $Pr$ as is shown in the inset of Fig. \ref{Figure_ste08_thermalBL}a. This effect is visible in RRB as shown in Fig. 5b of Ref. \cite{ste09d}. Here it is shown that for the non-rotating case $\lambda_u < \lambda_\theta^{sl}$ when $Pr\lesssim 1$. When the rotation rate is increased, i.e. $Ro$ is lowered, this transition shifts towards higher $Pr$ and for $Ro=0.1$ $\lambda_u < \lambda_\theta^{sl}$ when $Pr\lesssim 9$.

In summary, the laminar rotating BL theory explains the $Ro^{1/2}$ scaling of the kinetic BL thickness in RRB convection and the shift of the position where $\lambda_u=\lambda_\theta$ towards higher $Pr$ when the flow is dominated by rotational effects.

\section{Model for smooth onset in RRB convection}
\label{Sec4}

In this section we will introduce a model in the spirit of the GL approach in order to describe the smooth increase in the heat transfer as a function of $Ro$ that is observed for relatively low $Ra$ and weak background rotation. Since the GL theory assumes smooth transitions between different turbulent states the model is limited to the relatively low $Ra$ number regime, since a sharp onset as function of $Ro$ is found for $Ra\gtrsim 1\times10^8$. Furthermore, we assume that then the LSC, a basic ingredient of the GL model, is still present. In this simple model we neglect the influence of Ekman pumping, because it is a local effect that is rather insignificant at weak background rotation. This is supported by the results in Fig. \ref{Fig_SCL09_flow} and the EPAPS document of Ref. \cite{ste09}, where we find no evidence for Ekman pumping at weak background rotation. When strong rotation is applied Ekman pumping is the dominant effect and the validity regime of our model is thus restricted to weak background rotation. The basic idea of the model is to combine the effect of the LSC roll, which is implemented in the GL theory by the use of the laminar Prandtl-Blasius theory over an infinitely large plate, and the influence of rotation on the thermal BL.

Applying laminar BL theory requires that the viscous BL above the flat rotating plate does not brake down. This assumption will now be verified. The stability for the von K\'arm\'an flow has been studied theoretically and experimentally by Lingwood \cite{lin95,lin96,lin97}, showing that instability occurs at $Re \approx 510$ where $Re$ is defined as $Re=r\sqrt{\Omega/\nu}$ (with $r$ the distance to the rotation axis). Lingwood also pointed out that other experimental studies show the same transition point within a very narrow Reynolds number range $Re = 513 \pm 15$, see \cite{lin95} and references therein. More recent experiments show similar results \cite{col99,zou03}. For the Ekman case the instability occurs at $Re\approx200$, see \cite{lin97}. For the case under consideration it can be shown that $Re\lesssim 55$ (with $r=12.5~{\rm{cm}}$, $\Omega\approx 0.20~{\rm{rad/s}}$, and $\nu=1\times 10^6~{\rm{m}}^2/{\rm{s}}$) \cite{zho09b}. It can safely be conjectured that laminar BL theory can be applied as the estimated Reynolds number is an order of magnitude smaller than the critical Reynolds number. For further discussion on the stability of the rotational flow we refer to the classic Refs. \cite{cha81,bus70}.

We introduce $\lambda_{\theta R}=\overline{\lambda}_\theta^{sl}(Pr)\sqrt{\nu/\Omega}$, see $\overline{\lambda}_\theta^{sl}(Pr)$ in Fig. \ref{Figure_ste08_thermalBL}a, as the thermal BL thickness based on the background rotation and $\lambda_{\theta C}$ as the BL thickness based on the LSC roll. Furthermore, $\Gamma$ is the diameter-to-height aspect ratio of the RB cell and we set the radial length $r=(\Gamma L)/2$. Note that this is analogous to the length $L$ which is introduced in the GL theory for the length of the plate. Thus the Reynolds number based on the background rotation is

\begin{equation}
    Re_R=\frac{\Omega L^2 \Gamma^2}{4 \nu} \propto \frac{1}{Ro}~.
\end{equation}
To calculate $\lambda_{\theta R}$ we used $\nu = 1 \times 10^{-6} m^2/s$ (water) and $\Gamma=1$ and we set $\lambda_{\theta R}=\lambda_{\theta C}$ at $Ro =\infty$. The strength of the LSC roll is taken constant and $\lambda_{\theta C}$ is known from $\lambda_{\theta C}/L$ = $(2 Nu)^{-1}$.

We model the increase of $Nu$ as a crossover between a convection role dominated BL and a rotation dominated BL. Thus without rotation the BL thickness is determined by the LSC roll, i.e. $\lambda_{\theta}= \lambda_{\theta C}$, and when rotation becomes dominant the BL thickness is determined by the rotating BL thickness, i.e. $\lambda_{\theta}= \lambda_{\theta R}$. We now model the crossover between these two limiting cases as
\begin{equation} \label{Eq crossover}
    \frac{\lambda_{\theta}}{L}=\frac{\sqrt{Re_R}\lambda_{\theta R}+\alpha^* \sqrt{Re_C}\lambda_{\theta C}}{(\sqrt{Re_R} + \alpha^* \sqrt{Re_C})L}.
\end{equation}
Here the square root of Reynolds has been chosen since the dimensionless BL thicknesses scale with $1/\sqrt{Re}$. We rewrite the above equation in terms of $Ro$, using $Re_R \propto Re_C/Ro$,
\begin{equation} \label{Eq model}
    \frac{\lambda_\theta}{L} =\frac{\frac{1}{\alpha \sqrt{Ro}} \lambda_{\theta R} + \lambda_{\theta_C}}{\left(\frac{1}{\alpha\sqrt{Ro}}+1 \right)L}
\end{equation}
and we use the free parameter $\alpha$ to fit the model with the numerical data shown in Fig. \ref{Fig_SCL09_model}. It can be concluded that the presented model, based on the approach of the GL theory, indeed reflects the increase of $Nu$ (as compared to the case without rotation) observed at relatively low $Ra$ number ($Ra=4\times10^7$) when the LSC is still present. Furthermore, the thickness of the thermal BLs is also reflected correctly by the model. Note that the large value of the parameter $\alpha=55$ indicates that the influence of the rotation is rather weak before Ekman pumping sets in and it also explains that for higher $Ra$ ($Ra\gtrsim 1\times10^8$) no heat transfer enhancement is observed below onset. The sudden (instead of smooth) transition is then fully determined by the rotation rate where Ekman pumping sets in. This may be because at higher $Ra$ the thermal BL is already much thinner due to the stronger LSC and therefore the effect of weak rotation is not sufficient to result in a significant thinner thermal BL. When $Ro < Ro_c$ the model cannot be used, since Ekman pumping is dominant in this regime which is responsible for the strong increase observed in $Nu$ when $Ro < Ro_c$.

\section{Conclusions}
To summarize, we have studied the effect of rotation on the RB system at relatively low $Ra$ number, i.e. $Ra=4\times10^7$ by using DNS. We find a smooth increase of the heat transfer as function of the rotation rate when weak rotation is applied. To describe this heat transfer enhancement we have extended the GL theory to the rotating case by studying the influence of rotation on the scaling of the thermal BL thickness. It is based on a similar approach as in the laminar Prantl-Blasius BL theory over an infinitely large plate, as we analyzed the flow over an infinitely large rotating disk where the fluid at infinity is allowed to rotate. Just as in the Prantdl-Blasius BL theory we used a passive temperature field to calculate the characteristics of the thermal BL. It turns out that weak background rotation does not influence the scaling of the BL thickness in the high $Pr$ regime, because again $Pr^{-1/3}$ scaling is found. However, rotation does influence the scaling in the low $Pr$ regime where we find a scaling of $Pr^{-1}$ instead of $Pr^{-1/2}$ found in the Prandtl-Blasius BL theory. With our model for the thermal BL thickness, see Eq.~(\ref{Eq model}), we can explain the increased heat transfer observed in the relatively low $Ra$ number regime before the strength of the LSC decreases. The model neglects the effect of Ekman pumping as this effect is rather insignificant before the strength of the LSC decreases, i.e. the regime to which the model is applied. This means that the model cannot predict the heat transfer enhancement that is observed at moderate rotation rates where Ekman pumping is the dominant mechanism. The contrast between the smooth onset at $Ra=4\times10^7$ and the sharp onset at $Ra\gtrsim 1\times10^8$ is remarkable since only a small shift in the $Ra-Pr-Ro$ phase space is involved.\\

\emph{Acknowledgments}: We thank R. Verzicco for providing us with the numerical code and F. Fontenele Araujo, F. Busse, G.J.F. van Heijst, C. Sun, and L. van Wijngaarden for discussions. The work is sponsored by the Foundation for Fundamental Research on Matter (FOM) and the National Computing Facilities (NCF), both sponsored by NWO. The numerical calculations have been performed on the Huygens cluster of SARA in Amsterdam.

%
\begin{appendix}
\section{Ekman boundary layer theory}
\label{appA}
In this appendix the results obtained from the model with weak background rotation will be compared with analytic results obtained from Ekman BL theory \cite{gre90}, which uses a rotating reference frame. In the Ekman case the fluid at infinity is rotating at almost the same velocity as the disk, i.e. the limiting case of the model will be checked. We will indicate all quantities calculated in the rotating reference frame with an asterisk.

We will use the BEK model, presented in \cite{fal91,lin97,jas05}, to derive a similar ODE as in section \ref{Sec3} for the temperature advection equation in the rotating reference frame. In the BEK model the following self-similarity assumption for the axial velocity is proposed:

\begin{equation}
    \label{Eq assumption axial velocity BEK}
    w=\sqrt{\nu \Omega^{*}} Ro^* H^*(\xi),
\end{equation}
where $\xi= z \sqrt{\Omega^*/\nu}$. Here, $\Omega^{*}$ is a system rotation rate, and $Ro^*$ is a constant determined by the rotation rate. We call this still the Rossby number since it also represents a dimensionless inverse rotation. In the BEK model $Ro^*$, $\Delta \Omega$, and $\Omega^{*}$, respectively, are defined as
\begin{equation} \label{Eq Rossbystar}
    Ro^*=\frac{\Delta \Omega}{\Omega^{*}},
\end{equation}
\begin{equation} \label{Eq deltaOmega}
    \Delta \Omega = \Omega_F - \Omega_D,
\end{equation}
\begin{equation} \label{Eq Omegastar}
    \Omega^{*}=\frac{\Omega_F}{2-Ro^*}+\frac{\Omega_D}{2+Ro^*}.
\end{equation}
We obtain the following temperature advection equation in the BEK model
\begin{equation} \label{Eq ODE energy equation rotating reference frame}
    \widetilde{\theta}^{''} = Pr H^{*}(\xi) Ro^* \widetilde{\theta}'.
\end{equation}
From Eqs. (\ref{Eq Rossbystar})-(\ref{Eq Omegastar}) one obtains, after some algebra,
\begin{equation}
    s=\frac{\Omega_F}{\Omega_D}= \left[\frac{2+Ro^*-Ro^{*2}}{2-Ro^*-Ro^{*2}}\right].
\end{equation}
This means for the Ekman case ($s \approx 1$, i.e. $Ro^* \rightarrow 0 $) that $\Omega_F \simeq \Omega_D (1+Ro^*)$.
Thus when the fluid at infinity is rotating slower than the disk $Ro^*$ is negative. From now on we assume $Ro^*$ to be negative, i.e. $s<1$.

Using Ekman BL theory \cite{gre90} one can derive analytic solutions for the radial and tangential velocity profiles in the rotating reference frame. These analytical solutions read \cite{gre90}:
\begin{eqnarray}
    \label{Eq solution Ekman radial}
    u_E &=& - \Delta \Omega r e^{-\zeta} \sin \zeta =  r\Omega^* Ro^* F^*(\zeta)~,\\
    \label{Eq solution Ekman tangential}
    v_E &=& \Delta \Omega r(1 - e^{-\zeta} \cos \zeta) = r\Omega^* Ro^* G^*(\zeta)~,
\end{eqnarray}
with $\zeta$ as in (\ref{Eq dimensionless height}). With the analytic expression for the radial velocity (\ref{Eq solution Ekman radial}) and the continuity equation one obtains
\begin{eqnarray} \label{Eq solution Ekman axial}
    w_E & = & \Delta \Omega \sqrt{\frac{\nu}{\Omega_D}}\left(1 - e^{-\zeta} \left[\sin \zeta + \cos \zeta \right]\right)\nonumber\\
\nonumber\\
& = & \sqrt{\nu \Omega^*} Ro^* H^*(\zeta)~.
\end{eqnarray}
In the case $Ro^* \rightarrow 0$ it is found that $\Omega_D \approx \Omega^*$, thus $\xi \approx \zeta$. In particular, the expression for the axial flow reduces to $H^*(\xi)=\sqrt{\Omega^*/\Omega_D}(1-e^{-\xi} \left[\sin \xi + \cos \xi \right])$, where $\sqrt{\Omega^*/\Omega_D} \approx 1$. We find that the analytic expressions and the above numerical solutions are identical within numerical accuracy for the limiting case (Ekman solution, $s\approx1, Ro^* \rightarrow 0$). (Note that a coordinate transformation has to be applied as the Ekman solution is expressed in the corotating reference frame, whereas the numerical solution has been defined in the laboratory frame.)

Now we use this approach to calculate the BL characteristics for the Ekman layer. With the temperature advection equation (\ref{Eq ODE energy equation rotating reference frame}) and the analytic expression $H^*(\xi)$ for the axial velocity we determine the effective scaling exponent $\gamma$ in $\overline{\lambda}_\theta \sim Pr^{\gamma}$ as function of $Pr$ for $Ro^*=-10^{-3}$ and $Ro^*=-10^{-5}$. Fig. \ref{Figure_ste08_thermalBL} shows that when $Ro^*$ goes to zero the low $Pr$ regime ($\overline{\lambda}_\theta \gg \overline{\lambda}_u$) is extended to higher $Pr$, because the kinetic BL thickness becomes thinner and the axial velocity becomes smaller, i.e. the thermal BL becomes thicker.

Substitution of $H^*(\xi)$ in (\ref{Eq ODE energy equation rotating reference frame}) yields the following expression for the temperature gradient in the Ekman layer
\begin{equation} \label{Eq temperature gradient Ekman BL}
    \frac{d\widetilde{\theta}}{d\xi}=C_1 e^{\left[ \left(e^{-\xi}\cos \xi +\xi \right)Pr Ro^* \right]},
\end{equation}
where $C_1$ is a constant of integration which does not depend on $\xi$. To determine the constant $C_1$, and thereby $Nu$, one needs to integrate Eq.\ (\ref{Eq temperature gradient Ekman BL}). An analytic result can be derived by substituting $\cos \xi = (e^{i\xi}+e^{-i\xi})/2$, $A=PrRo^*$, $z=A\xi$, $B=(i-1)/A$ (and $\overline{B}$ the complex conjugate of $B$), and evaluate the integral

 \begin{equation}  \label{Eq AB}
 \widetilde{\theta}(z) = \frac{C_1}{A}\int e^z e^{\frac{1}{2}Ae^{Bz}} e^{\frac{1}{2}Ae^{\overline{B}z}}dz + C_2~.
 \end{equation}

The integration constants $C_1$ and $C_2$ are determined by the boundary conditions $\widetilde{\theta}(\xi=0)=1$ and $\widetilde{\theta}(\xi\rightarrow \infty)=0$: $C_2=0$ and, for small $A$,

 \begin{equation} \label{Eq coeff1}
 \frac{1}{C_1}\approx \frac{1}{A}-\frac{1}{2}A-\frac{3}{16}A^2+{\mathcal{O}}(A^3)~.
 \end{equation}

The thermal BL thickness scales according to $\overline{\lambda}_\theta\propto e^{-A}/C_1$. With (\ref{Eq coeff1}) we immediately see that in the small Prandtl number limit: $\overline{\lambda}_\theta\propto Pr^{-1}$. Comparison with Fig. \ref{Figure_ste08_thermalBL}b reveals that the analytic results are in good agreement with the numerical data for $\overline{\lambda}_\theta$ represented by the solid lines. Moreover, it predicts the scaling for $Ro^*=-1$ (the von K\'arm\'an case, thus far outside the regime of applicability of the Ekman analysis) surprisingly well, see Fig. \ref{Figure_ste08_thermalBL}b.

\end{appendix}
\bibliographystyle{prsty_withtitle}


\end{document}